\documentclass[english,aps, pre, superscriptaddress, twocolumn]{revtex4-2}
\usepackage[T1]{fontenc}
\usepackage[latin9]{inputenc}
\usepackage{color}
\usepackage{babel}
\usepackage{amsmath}
\usepackage{amssymb}
\usepackage{graphicx}
\usepackage[unicode=true,pdfusetitle,
 bookmarks=true,bookmarksnumbered=false,bookmarksopen=false,
 breaklinks=false,pdfborder={0 0 0},pdfborderstyle={},backref=false,colorlinks=true]
 {hyperref}
\hypersetup{
 linkcolor=mycolor, citecolor=mycolor, urlcolor=mycolor}

\makeatletter

\usepackage{color}
\usepackage{babel}
\usepackage{breakurl}
\usepackage{orcidlink}

\renewcommand{\fnum@figure}{FIG.~\thefigure}

\definecolor{mycolor}{rgb}{0.0, 0.0, 0.55}

\@ifundefined{showcaptionsetup}{}{
 \PassOptionsToPackage{caption=false}{subfig}}

\usepackage[caption=false]{subfig} 

\makeatother

\begin{document}
\title{Multiphase nonlinear electron plasma waves}
\author{Vadim R. Munirov\orcidlink{0000-0001-6711-1272}}
\email{vmunirov@berkeley.edu}

\affiliation{Department of Physics, University of California, Berkeley, California
94720, USA}
\author{Lazar Friedland\orcidlink{0000-0002-3603-6908}}
\affiliation{Racah Institute of Physics, Hebrew University of Jerusalem, Jerusalem
91904, Israel}
\author{Jonathan S. Wurtele\orcidlink{0000-0001-8401-0297}}
\affiliation{Department of Physics, University of California, Berkeley, California
94720, USA}
\date{PRE: Received 28 May 2022; accepted 14 October 2022, published 3 November 2022}
\begin{abstract}
We present a method for constructing multiphase excitations in the
generally nonintegrable system of warm fluid equations describing
plasma oscillations. It is based on autoresonant excitation of nonlinear
electron plasma waves by phase locking with small amplitude chirped-frequency
ponderomotive drives. We demonstrate the excitation of these multiphase
waves by performing fully nonlinear numerical simulations of the fluid
equations. We develop a simplified model based on a weakly nonlinear
analytical theory by applying Whitham\textquoteright s averaged Lagrangian
procedure. The simplified model predictions are in good agreement
with the results from the warm fluid simulations. Such autoresonantly
excited multiphase waves form coherent quasicrystalline structures,
which can potentially be used as plasma photonic or accelerating devices.
Finally, we discuss the laser parameters required for the autoresonant
excitation of nonlinear waves in a plasma.
\end{abstract}
\maketitle

\section{\label{sec_I}Introduction}

The electron plasma wave is perhaps the most studied collective oscillation
in a plasma, yet the nonlinear behavior of the electron plasma wave,
including fluid and kinetic effects, remains an active topic of research
\citep{Einaudi1969,Manheimer1969,Coffey1971,Morales1972,Dewar1972,Dewar1972b,Cohen1977,Lindberg2007,Dodin2011,Berger2013,Benisiti2016,Dubin2015,Liu2015,Tacu2022a,Benisti2022,Winjum2007}
even after decades of study. Nonlinear effects in electron plasma
waves are important for many applications of laser-plasma interactions,
including ultra-high gradient accelerators \citep{Tajima1979}, inertial
confinement fusion (ICF) \citep{Kruer1995}, and photonics for extremely
intense laser pulses~\citep{Milchberg2019}.

In most applications, the electron plasma wave is controlled by the
ponderomotive force of one or more lasers. For example, in plasma
photonics, complex density structures are envisioned for transient
plasma gratings \citep{Lehmann2016}, holographic gratings \citep{Edwards2022},
and polarizers \citep{Michel2014,Turnbull2016,Lehmann2018,Kur2021}.
Resonant plasma instabilities, and the concomitant density modulations,
lead to energy transfer between laser pulses. In the Raman backscatter
amplifier, energy in a long laser pulse is transferred to a counterpropagating
short pulse \citep{Malkin1999}. Crossbeam energy transfer \citep{Michel2009}
makes use of a resonance with an ion acoustic wave and is routinely
utilized to control asymmetries in target illumination at the National
Ignition Facility (NIF). 

Autoresonance is a phenomenon of nonlinear science that has ample
applications in plasma, astro-, and atomic physics \citep{Fajans2001,Friedland2009}.
The basic idea of autoresonance lies in the ability of a nonlinear
system to remain in resonance by phase locking (synchronization) with
external drives with adiabatically varying parameters. It has been
proposed as a method to create a large amplitude traveling plasma
wave using two copropagating lasers with a chirped frequency mismatch
\citep{Lindberg2004} that passes through the linear electron plasma
wave frequency. In spatial autoresonance, two constant frequency lasers
propagate parallel to a plasma density gradient \citep{Yaakobi2008}
with the linear resonance at a specific location in the plasma. In
autoresonance, the nonlinear oscillator maintains synchronism with
a chirped-frequency drive so long as a threshold condition, which
relates the chirp rate to the drive amplitude, is satisfied. Autoresonance
can be used as a method to excite large amplitude traveling ion acoustic
waves \citep{Friedland2014,Friedland2017}.

Autoresonant excitation of electron plasma and ion acoustic waves
is not limited to traveling waves. It was shown in Ref.~\citep{Friedland2019}
that a large amplitude standing ion acoustic wave can be formed using
two counterpropagating ponderomotive drives with a chirped frequency
difference. This standing wave comprises a particular nonlinear two-phase
ion acoustic wave structure, wherein each locked phase corresponds
to one of the counterpropagating traveling drives. Large amplitude
standing electron plasma waves can be created with autoresonant drives~\citep{Friedland2020}.

It has been shown, using both theory and numerical simulations, that
one can use autoresonance to construct multiphase solutions for integrable
systems, such as the Korteweg\textendash de Vries (KdV) equation \citep{Friedland2003},
the Toda lattice \citep{Khasin2003}, the nonlinear Schr\"{o}dinger
equation \citep{Friedland2005}, and the sine-Gordon equation \citep{Shagalov2009}.
Multiphase nonlinear waves are significantly more difficult to analyze
theoretically than traveling waves, which can be described by a single
phase. The theory for autoresonant wave excitation, nevertheless,
has been extended to two-phase nonlinear waves. In a recent publication~\citep{Munirov2022a},
we demonstrated how autoresonance can be used to create two-phase
solutions in the generally nonintegrable system of equations describing
ion acoustic waves. This extends the earlier work where the autoresonant
excitation was analyzed for nonlinear single-phase \citep{Friedland2014,Friedland2017}
and standing \citep{Friedland2019} ion acoustic waves. In Ref.~\citep{Friedland2020},
it was shown that autoresonance can be used to excite large amplitude
standing electron plasma waves, which can be regarded as a particular
case of a more general two-phase solution. In this paper we will show
that the system describing plasma waves indeed exhibits nonlinear
two-phase solutions that are characteristic of integrable partial
differential equations. Similar to the case of ion acoustic waves~\citep{Munirov2022a},
space-time quasicrystalline structures formed by the autoresonantly
excited multiphase nonlinear plasma waves can potentially be used
as plasma photonic or, perhaps, even, as specialized accelerating
structures.

\noindent 
\begin{figure}
\textbf{(a)}

\includegraphics[width=1\columnwidth]{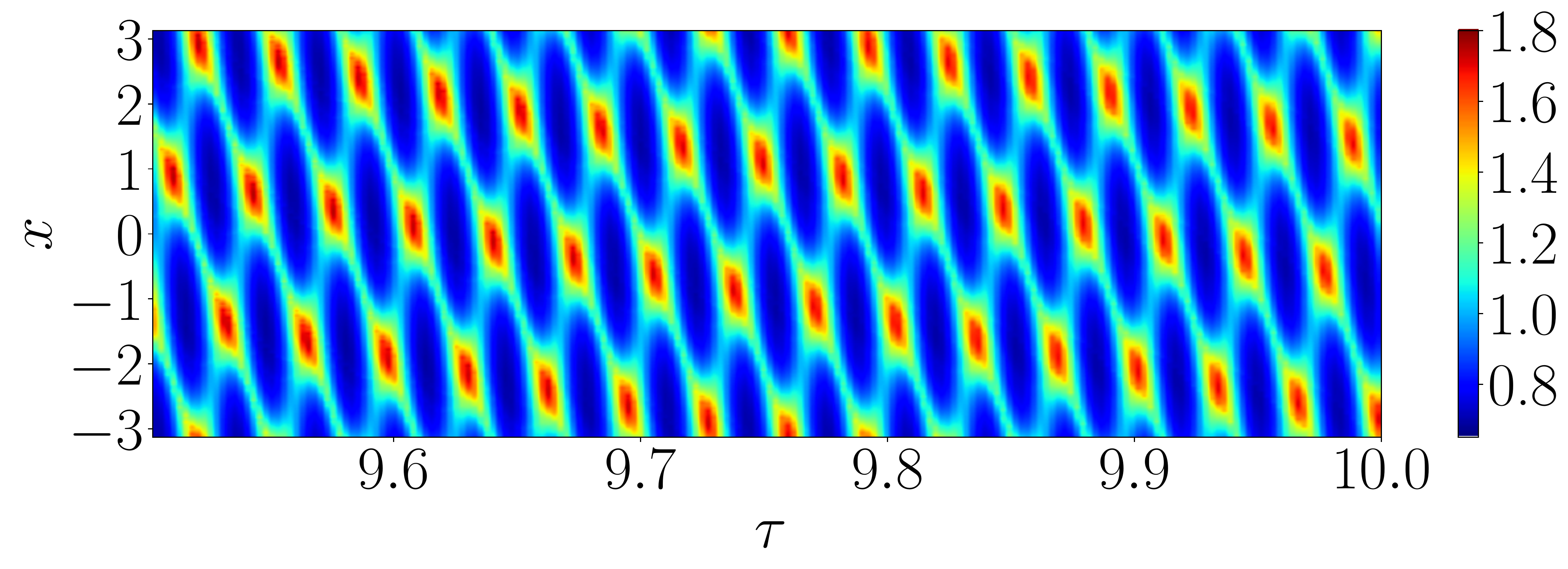}

\textbf{(b)}

\includegraphics[width=1\columnwidth]{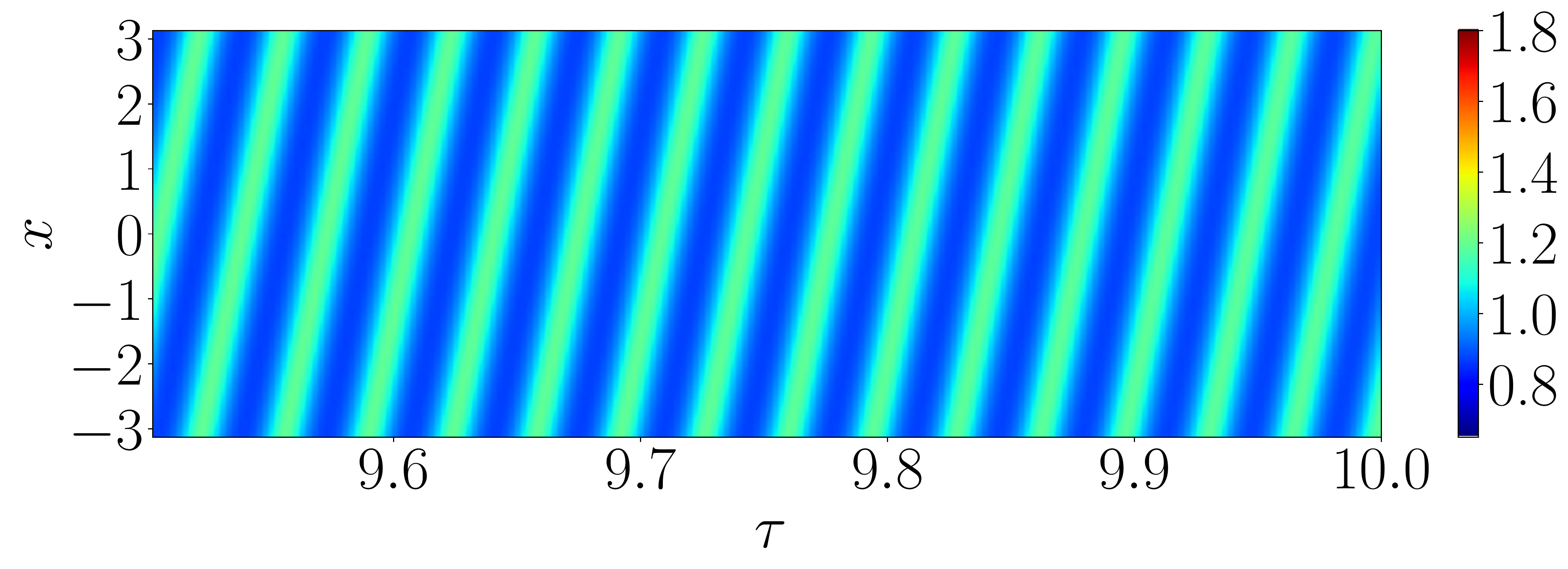}

\textbf{(c)}

\includegraphics[width=1\columnwidth]{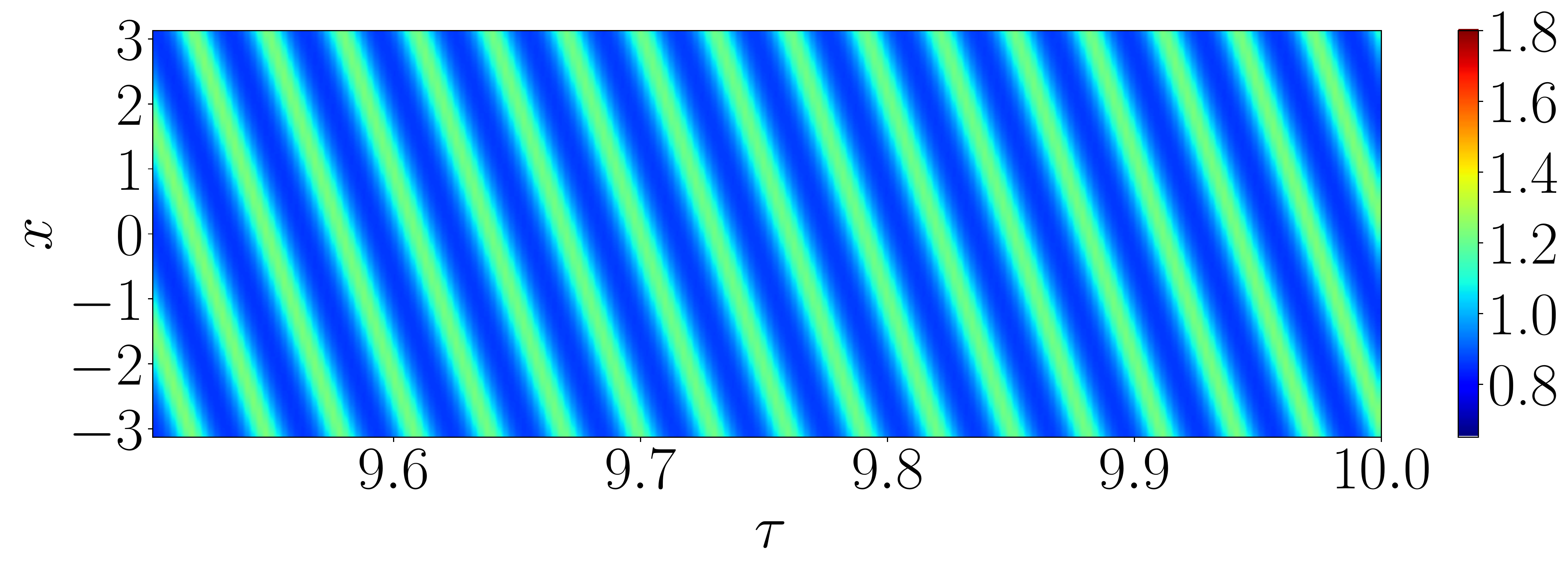}

\caption{\label{fig_n_nl}The colormap of the electron density $n_{e}\left(x,\tau\right)$
as a function of slow time $\tau=\sqrt{\alpha_{1}}t$ and coordinate
$x$ obtained by solving the fully nonlinear equations~(\ref{eq:sigma_xt})\textendash (\ref{eq:phi_xx_2}).
(a)~Two-phase autoresonant plasma wave excited by two driving counterpropagating
traveling waves with $k_{1}=1$ and $k_{2}=-2$. (b)~Single-phase
autoresonant plasma wave excited solely by the first driving component
with $k_{1}=1$. (c)~Single-phase autoresonant plasma wave excited
solely by the second driving component with $k_{2}=-2$.}
\end{figure}
\begin{figure}
\textbf{(a)}

\includegraphics[width=1\columnwidth]{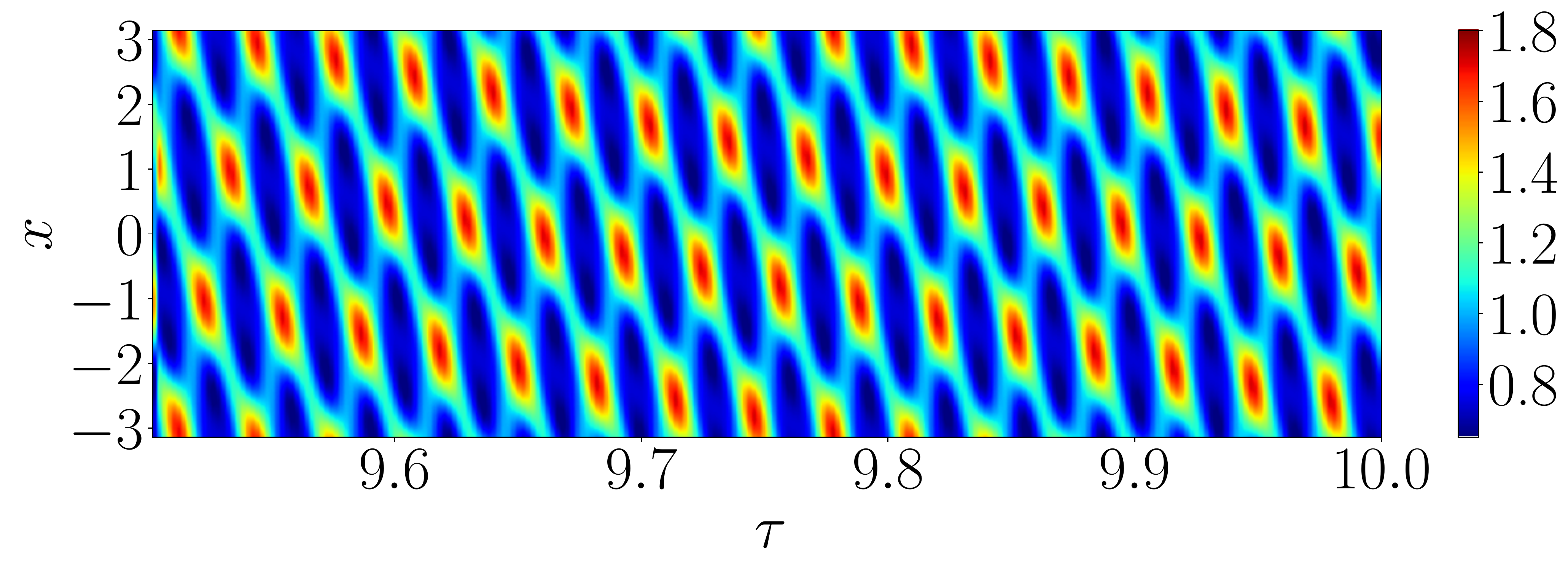}

\textbf{(b)}

\includegraphics[width=1\columnwidth]{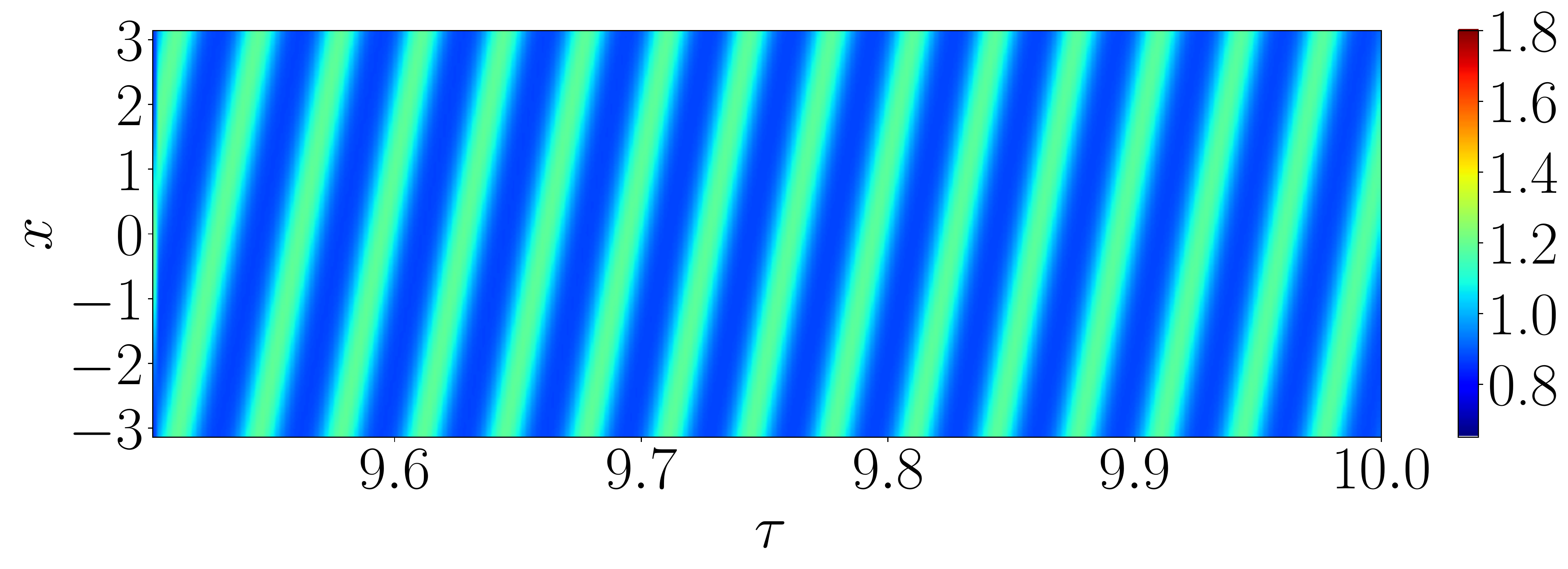}

\textbf{(c)}

\includegraphics[width=1\columnwidth]{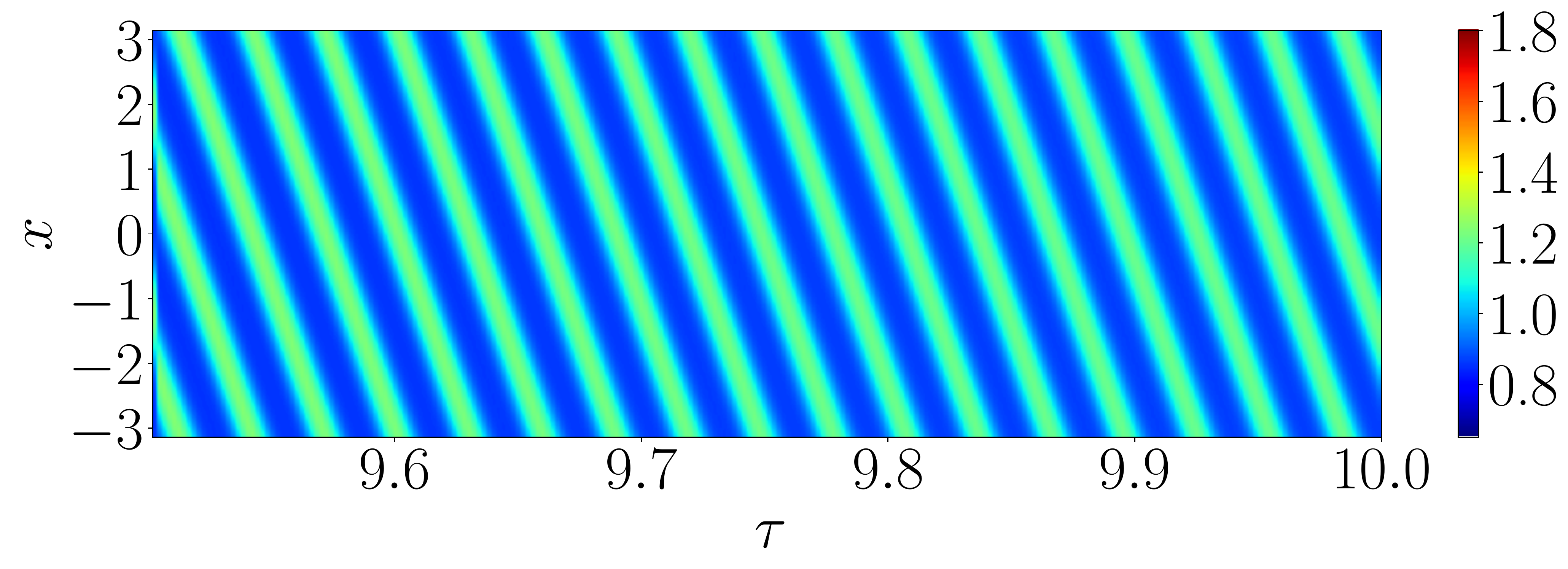}

\caption{\label{fig_n_weakly}The colormap of the electron density $n_{e}\left(x,\tau\right)$
as a function of slow time $\tau=\sqrt{\alpha_{1}}t$ and coordinate
$x$ obtained by solving the weakly nonlinear equations~(\ref{eq:dI1_dt})\textendash (\ref{eq:dPHI2_dt}).
(a)~Two-phase autoresonant plasma wave excited by two driving counterpropagating
traveling waves with $k_{1}=1$ and $k_{2}=-2$. (b)~Single-phase
autoresonant plasma wave excited solely by the first driving component
with $k_{1}=1$. (c)~Single-phase autoresonant plasma wave excited
solely by the second driving component with $k_{2}=-2$.}
\end{figure}

This paper is organized as follows. In Sec.~\ref{sec_II}, we present
a warm fluid model of partial differential equations describing electron
plasma waves, and we demonstrate, through fully nonlinear numerical
simulations, that it supports a two-phase solution. In Sec.~\ref{sec_III},
we apply Whitham\textquoteright s averaged variational principle \citep{Whitham1965,Whitham1999}
to the Lagrangian formulation of the fluid equations and develop an
analytical weakly nonlinear theory in the form of a system of coupled
ordinary differential equations; this system is shown to yield a good
approximation of the fully nonlinear model. In Sec.~\ref{sec_IV},
we estimate the laser pulse intensity and duration required for autoresonant
excitation of nonlinear plasma and ion acoustic waves. Finally, a
summary and concluding remarks are given in Sec.~\ref{sec_Conclusions}.

\section{\label{sec_II}Numerical study of the excitation of multiphase nonlinear
plasma waves}

A warm fluid model of electron plasma waves is constituted by a system
of continuity, momentum, and Poisson's equations:

\begin{gather}
\sigma_{xt}+\left[\left(1+\sigma_{x}\right)\psi_{x}\right]_{x}=0,\label{eq:sigma_xt}\\
\psi_{xt}+\psi_{x}\psi_{xx}=\left(\varphi+\varphi_{d}\right)_{x}-\Delta^{2}\left(1+\sigma_{x}\right)\sigma_{xx},\label{eq:psi_xt}\\
\varphi_{xx}=\kappa^{2}\varphi+\sigma_{x}.\label{eq:phi_xx_2}
\end{gather}

Here we introduced potentials $\sigma$ and $\psi$, which are defined
through $n_{e}=1+\sigma_{x}$, $v=\psi_{x}$, where $n_{e}$ is the
electron density and $v$ is the fluid velocity. Other variables are
the electric potential $\varphi$ and the driving potential~$\varphi_{d}$.
Parameter $\kappa$ is the effective screening parameter (see Refs.~\citep{Dubin2015,Friedland2020}),
while $\Delta^{2}=3u_{th}^{2}$, where $u_{th}$ is the electron thermal
velocity which is assumed constant. All variables are dimensionless;
specifically, the time is measured in terms of the inverse plasma
frequency $\omega_{p}^{-1}$, the distance is normalized to $k^{-1}$,
where $k$ is the typical wave vector, the plasma density is normalized
to the unperturbed plasma density, and the electric and driving potentials
are normalized to~$m_{e}\omega_{p}^{2}/ek^{2}$.

\noindent 
\begin{figure}
\textbf{(a)}

\includegraphics[width=1\columnwidth]{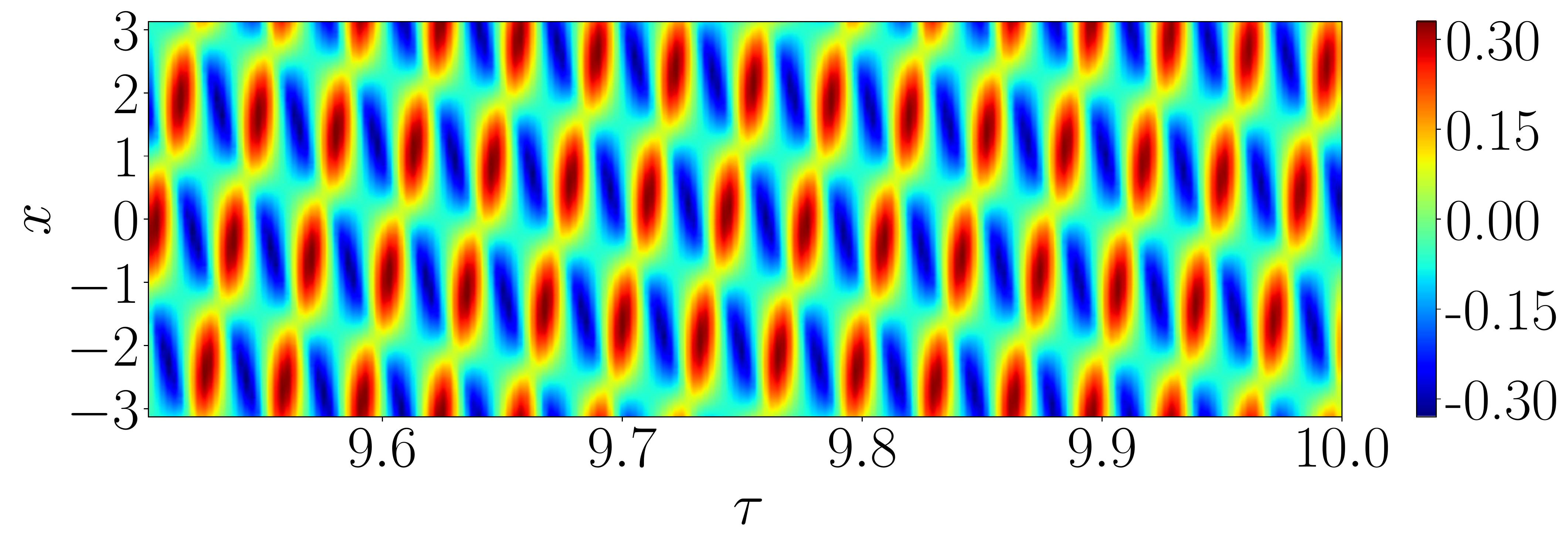}

\textbf{(b)}

\includegraphics[width=1\columnwidth]{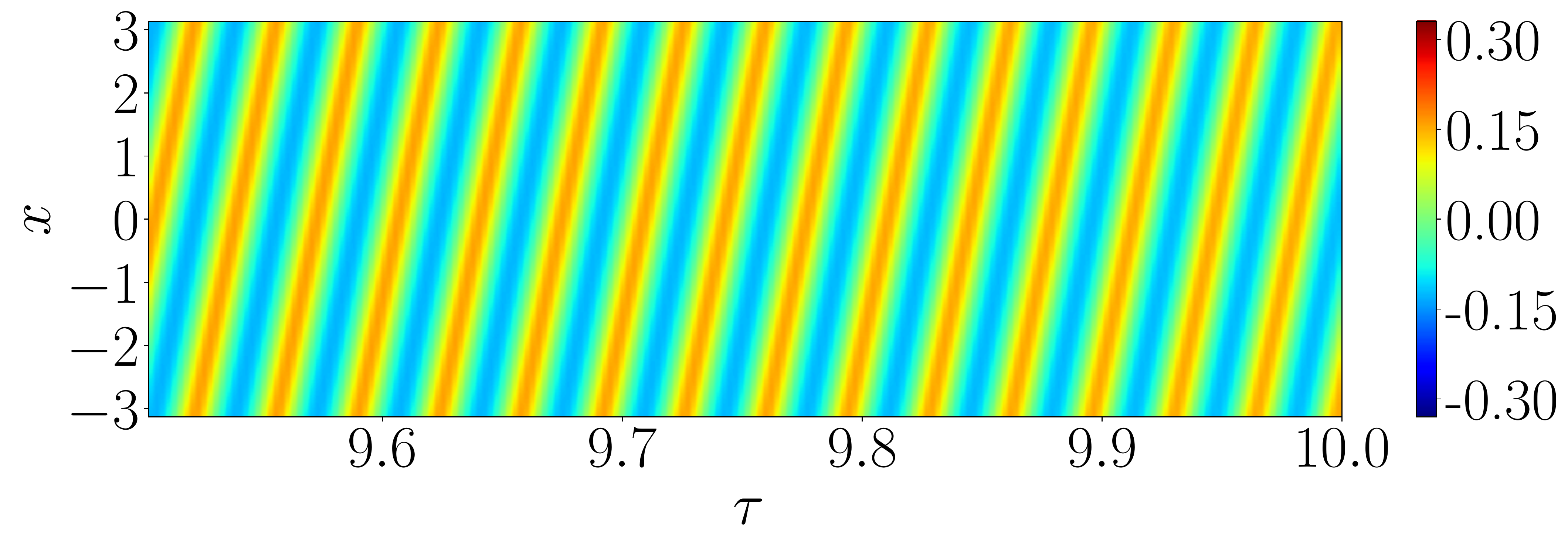}

\textbf{(c)}

\includegraphics[width=1\columnwidth]{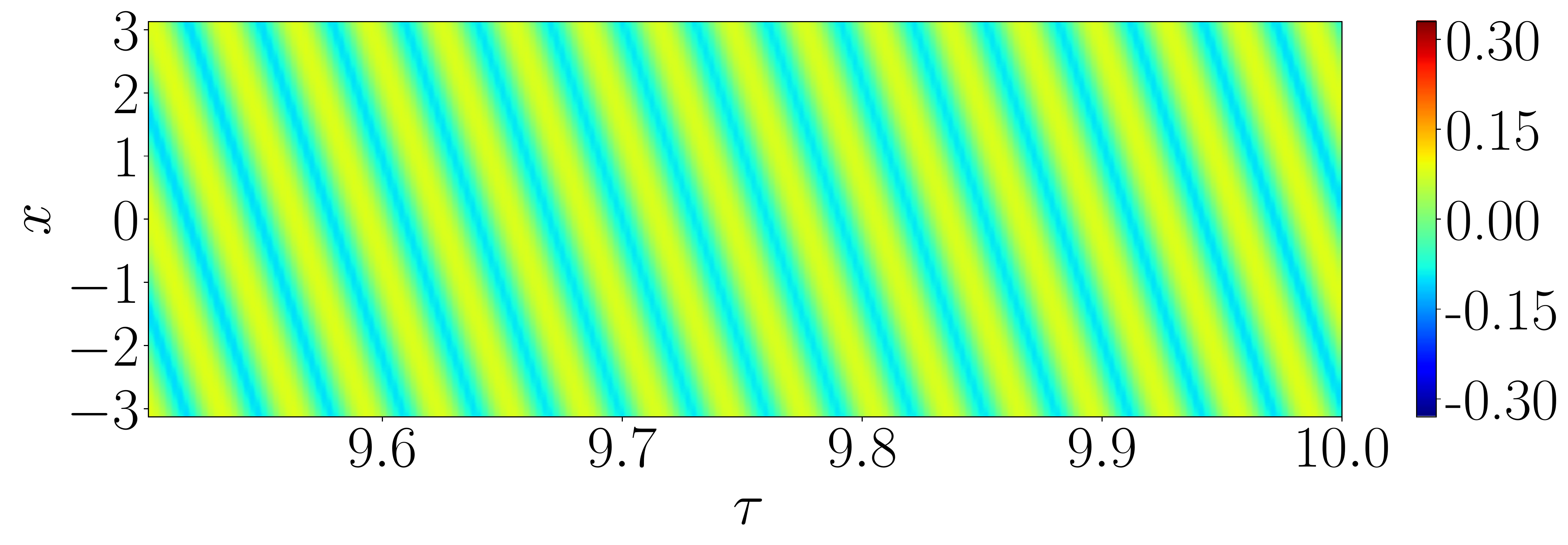}

\caption{\label{fig_vel_nl}The colormap of the electron fluid velocity $v\left(x,\tau\right)$
as a function of slow time $\tau=\sqrt{\alpha_{1}}t$ and coordinate
$x$ obtained by solving the fully nonlinear equations~(\ref{eq:sigma_xt})\textendash (\ref{eq:phi_xx_2}).
(a)~Two-phase autoresonant plasma wave excited by two driving counterpropagating
traveling waves with $k_{1}=1$ and $k_{2}=-2$. (b)~Single-phase
autoresonant plasma wave excited solely by the first driving component
with $k_{1}=1$. (c)~Single-phase autoresonant plasma wave excited
solely by the second driving component with $k_{2}=-2$.}
\end{figure}
\begin{figure}
\textbf{(a)}

\includegraphics[width=1\columnwidth]{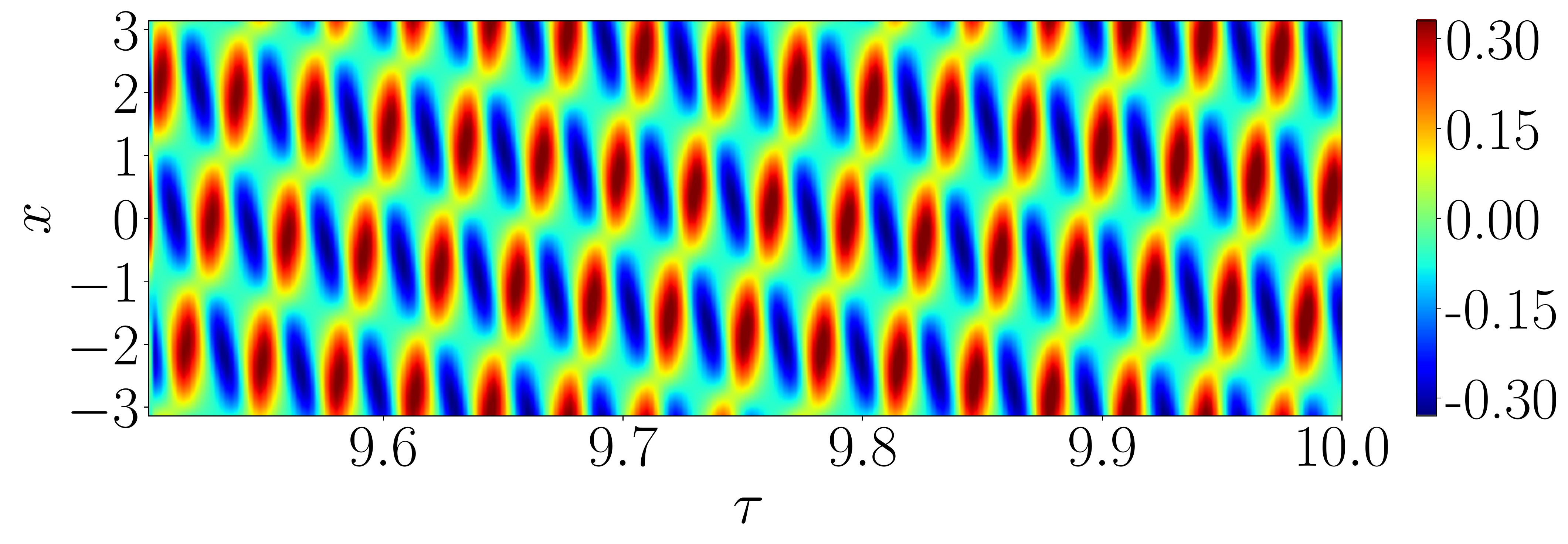}

\textbf{(b)}

\includegraphics[width=1\columnwidth]{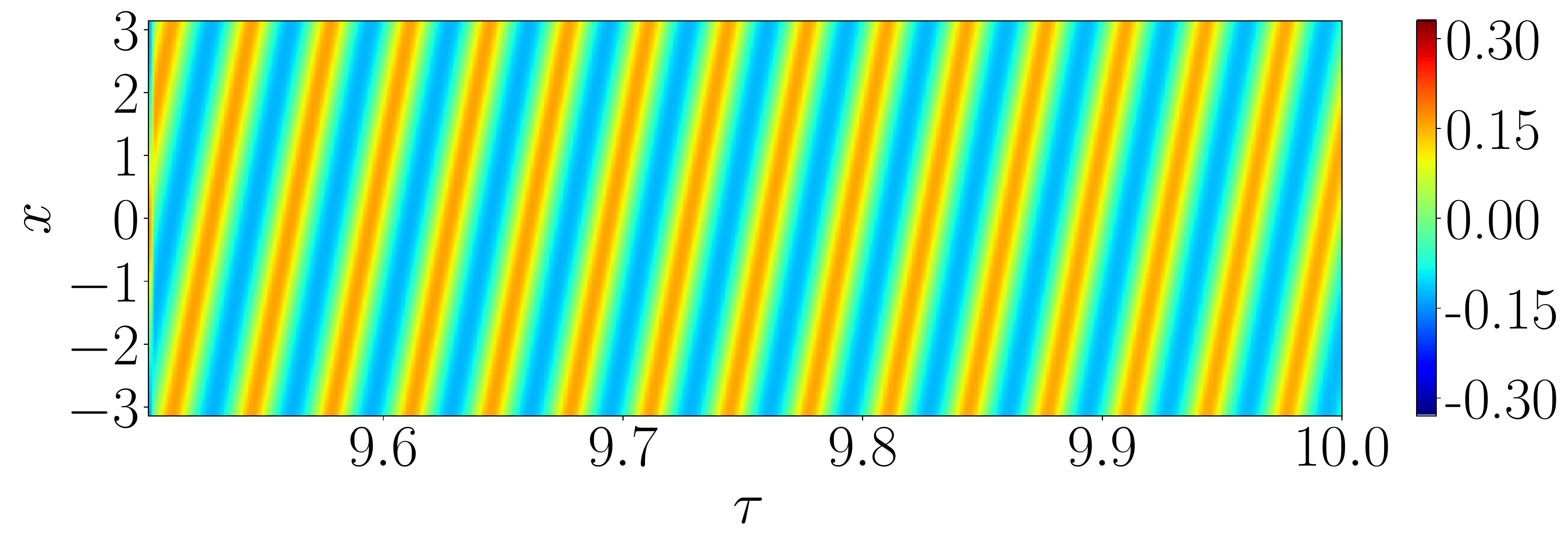}

\textbf{(c)}

\includegraphics[width=1\columnwidth]{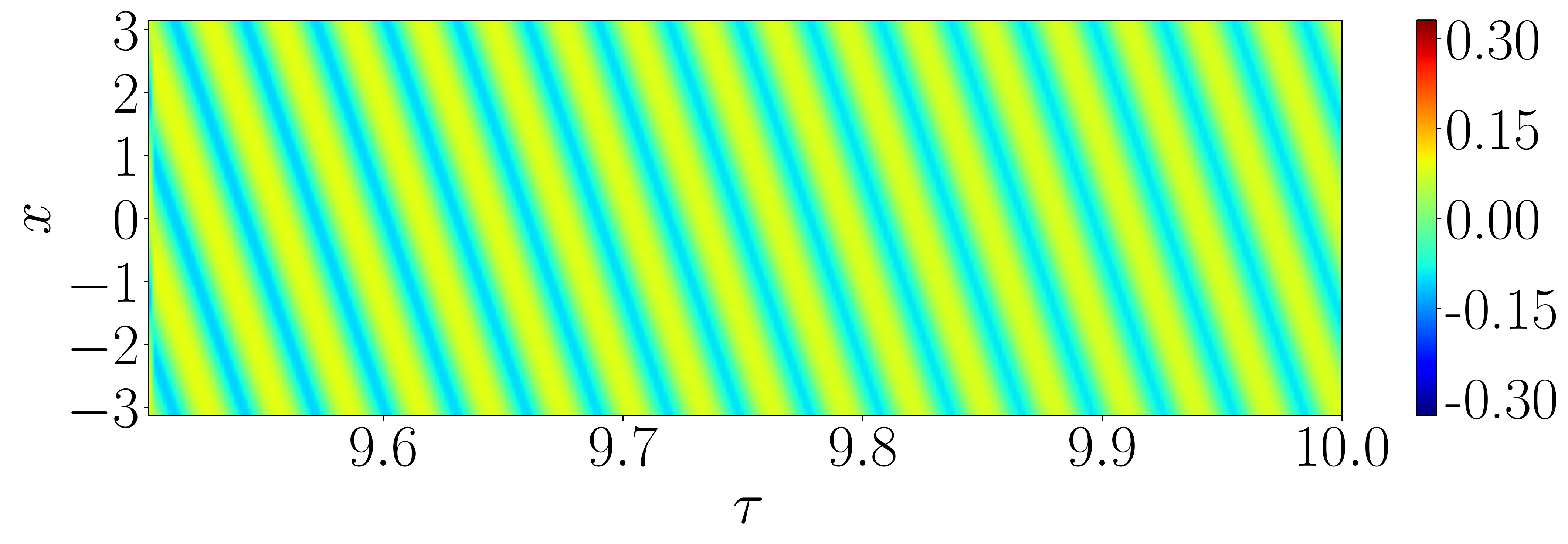}

\caption{\label{fig_vel_weakly}The colormap of the electron fluid velocity
$v\left(x,\tau\right)$ as a function of slow time $\tau=\sqrt{\alpha_{1}}t$
and coordinate $x$ obtained by solving the weakly nonlinear equations~(\ref{eq:dI1_dt})\textendash (\ref{eq:dPHI2_dt}).
(a)~Two-phase autoresonant plasma wave excited by two driving counterpropagating
traveling waves with $k_{1}=1$ and $k_{2}=-2$. (b)~Single-phase
autoresonant plasma wave excited solely by the first driving component
with $k_{1}=1$. (c)~Single-phase autoresonant plasma wave excited
solely by the second driving component with $k_{2}=-2$.}
\end{figure}

We consider the driving term consisting of the two small amplitude
traveling wave ponderomotive drives:
\begin{equation}
\varphi_{d}=\varepsilon_{1}\cos\left(\theta_{d,1}\right)+\varepsilon_{2}\cos\left(\theta_{d,2}\right),\label{eq:phi_d}
\end{equation}

\noindent where $\theta_{d,i}=k_{i}x-\int\omega_{d,i}\left(t\right)dt$
($i=1,2$) are driving phases with wave vectors $k_{i}$ and slowly
varying driving frequencies $\omega_{d,i}\left(t\right)=-d\theta_{d,i}/dt$.

To illustrate autoresonant excitation of single- and multiphase plasma
waves, let us consider a representative example of two driving counterpropagating
traveling waves with wave vectors $k_{1}=1$ and $k_{2}=-2$. We choose
their chirped driving frequencies as

\begin{equation}
\omega_{d,i}=\begin{cases}
\omega_{p,i}+\alpha_{i}t, & t\leq0,\\
\omega_{p,i}+\alpha_{i}T_{i}\arctan\left(\frac{t}{T_{i}}\right), & t>0,
\end{cases}\label{eq:w_d}
\end{equation}

\noindent where $T_{i}=2\Delta\omega_{i}/\pi\alpha_{i}$ ($i=1,2$),
$\alpha_{1}=\alpha_{2}=2.5\times10^{-5}$, $\Delta\omega_{1}=\Delta\omega_{2}=0.008$
and $\omega_{p,i}$ ($i=1,2$) are the frequencies given by the linear
plasma wave dispersion relation:

\begin{equation}
\omega_{p,i}\left(k_{i}\right)=\sqrt{\frac{1}{1+\frac{\kappa^{2}}{k_{i}^{2}}}+\Delta^{2}k_{i}^{2}}.\label{eq:w_p}
\end{equation}

We also gradually increase the driving amplitudes as $\varepsilon_{i}=\bar{\varepsilon}_{i}\left[0.5+\arctan\left(t\sqrt{\alpha_{i}}/10\right)/\pi\right]$,
$\bar{\varepsilon}_{i}=2\times10^{-3}$ ($i=1,2$). We take the electron
thermal velocity and the effective screening parameter to be $u_{th}=0.1$
and $\kappa=0.5$, respectively. We use the wave vector of the first
drive $k_{1}$ as the typical wave vector $k$ (hence, $k_{1}=1$).

The system of nonlinear partial differential equations~(\ref{eq:sigma_xt})\textendash (\ref{eq:phi_xx_2})
can be solved numerically. To do this we use a pseudospectral method
\citep{Canuto1987} in space and the fourth-order Runge\textendash Kutta
method for the time advancement, similar to the procedure employed
in Refs.~\citep{Friedland2017,Friedland2019,Friedland2020,Munirov2022a}.
We run simulations from $\tau=-10$ and stop the driving at $\tau=10$,
where $\tau=\sqrt{\alpha_{1}}t$ is a slow time variable. The results
of the fully nonlinear numerical simulations with the parameters specified
above are presented in Figs.~\ref{fig_n_nl}, \ref{fig_vel_nl},
and~\ref{fig_max_u_vel}.

Figure~\ref{fig_n_nl} shows a colormap of the electron density $n_{e}\left(x,\tau\right)$
as a function of slow time $\tau$ and coordinate $x$ for $\tau$
between $\tau=9.5$ and $\tau=10$. Figure~\hyperref[fig_n_nl]{\ref{fig_n_nl}(a)}
shows a two-phase nonlinear electron plasma wave excited by two small
amplitude chirped-frequency traveling waves with $k_{1}=1$ and $k_{2}=-2$.
Figure~\hyperref[fig_n_nl]{\ref{fig_n_nl}(b)} shows a single-phase
plasma wave autoresonantly excited only by the first chirped-frequency
traveling wave drive with $k_{1}=1$, using the same parameters in
the drive as in Fig.~\hyperref[fig_n_nl]{\ref{fig_n_nl}(a)} but
with vanishing $\varepsilon_{2}$, while Fig.~\hyperref[fig_n_nl]{\ref{fig_n_nl}(c)}
shows a single-phase plasma wave autoresonantly excited by the chirped-frequency
traveling wave drive with $k_{2}=-2$, using the same parameters in
the drive as in Fig.~\hyperref[fig_n_nl]{\ref{fig_n_nl}(a)} but
with vanishing $\varepsilon_{1}$. 

Figure~\ref{fig_vel_nl} is identical to Fig.~\ref{fig_n_nl} but
shows a colormap of the fluid velocity $v\left(x,\tau\right)$ as
a function of slow time $\tau$ and coordinate $x$ instead. We can
clearly see from Figs.~\ref{fig_n_nl} and~\ref{fig_vel_nl} that
a highly nonlinear large amplitude $\left(\delta n_{e}/n_{e}\sim1\right)$
two-phase, quasiperiodic in space and time, structure is excited by
the drives. We can also see that the directions of the phase velocities
of the single-phase waves {[}see Figs.~\hyperref[fig_n_nl]{\ref{fig_n_nl}(b)}
and~\hyperref[fig_n_nl]{\ref{fig_n_nl}(c)}{]} correspond to the
characteristic directions seen in the two-phase solution {[}see Fig.~\hyperref[fig_n_nl]{\ref{fig_n_nl}(a)}{]}.
We also note that, as in the case of ion acoustic waves \citep{Munirov2022a},
the nonlinear structures persist even after we turn off the small
amplitude drives (not shown in the figures).

Figure~\ref{fig_max_u_vel} shows the maximum value over $x$ of
the electron density $n_{e}\left(x,\tau\right)$ {[}Fig.~\hyperref[fig_max_u_vel]{\ref{fig_max_u_vel}(a)}{]}
and of the fluid velocity $v\left(x,\tau\right)$ {[}Fig.~\hyperref[fig_max_u_vel]{\ref{fig_max_u_vel}(b)}{]}
vs slow time $\tau=\sqrt{\alpha_{1}}t$ from the start of the simulation
at $\tau=-10$ to $\tau=10$. We can see that the system passes the
linear plasma resonance at $\tau=0$ and then the amplitudes of both
waves rapidly increase, reaching large values. This happens because,
due to nonlinear effects, the waves alter their amplitudes in a way
that lets them stay phase-locked with the external drives, allowing
the continuous transfer of energy from the drives to the excitations.

To better understand the nature of the double autoresonance, and to
have a tool to select the appropriate parameters required to establish
the autoresonance, we need to develop a theory. To that end, in the
next section, we will formulate the problem in the Lagrangian language
and then use Whitham's averaged variational principle \citep{Whitham1965,Whitham1999}
to obtain the simplified weakly nonlinear equations describing the
evolution of the system.

\section{\label{sec_III}Weakly nonlinear theory and Whitham's variational
method}

The system of nonlinear equations~(\ref{eq:sigma_xt})\textendash (\ref{eq:phi_xx_2})
can be described using the Lagrangian formalism. Indeed, one can check
that Eqs.~(\ref{eq:sigma_xt})\textendash (\ref{eq:phi_xx_2}) are,
in fact, the Euler\textendash Lagrange equations of the form

\begin{gather}
\frac{\partial L}{\partial\sigma}-\frac{\partial}{\partial t}\frac{\partial L}{\partial\sigma_{t}}-\frac{\partial}{\partial x}\frac{\partial L}{\partial\sigma_{x}}=0,\\
\frac{\partial L}{\partial\psi}-\frac{\partial}{\partial t}\frac{\partial L}{\partial\psi_{t}}-\frac{\partial}{\partial x}\frac{\partial L}{\partial\psi_{x}}=0,\\
\frac{\partial L}{\partial\varphi}-\frac{\partial}{\partial t}\frac{\partial L}{\partial\varphi_{t}}-\frac{\partial}{\partial x}\frac{\partial L}{\partial\varphi_{x}}=0,
\end{gather}

\noindent which emerge from the following Lagrangian density:

\begin{multline}
L=\frac{1}{2}\varphi_{x}^{2}+\frac{1}{2}\kappa^{2}\varphi^{2}-\frac{1}{2}\left(\psi_{t}\sigma_{x}+\psi_{x}\sigma_{t}\right)-\frac{1}{2}\psi_{x}^{2}\left(1+\sigma_{x}\right)\\
-\frac{1}{2}\Delta^{2}\sigma_{x}^{2}\left(1+\frac{1}{3}\sigma_{x}\right)+\sigma_{x}\left(\varphi+\varphi_{d}\right).\label{eq:Lagrangian}
\end{multline}

Since we use slowly varying driving frequencies, it is appropriate
to exploit a natural separation into slow and fast dynamics. We now
proceed to derive equations describing the slow evolution of the waves
in space and time. Whitham \citep{Whitham1965,Whitham1999} demonstrated
how to obtain such equations on the basis of the Lagrangian formalism.
In this section, we will employ Whitham's averaged Lagrangian method
to obtain weakly nonlinear equations describing the evolution of the
slow variables. We will closely follow the procedure that we used
to study multiphase ion acoustic waves described in Ref.~\citep{Munirov2022a}.

Let us first examine the linear stage of the evolution. If we start
from the equilibrium solution ($n_{e}=1$, $v=0$, $\varphi=0$),
then it is straightforward to show that during the linear stage the
solutions are given by

\begin{gather}
\sigma=\tilde{A}_{10}\sin\left(\theta_{1}\right)+\tilde{A}_{01}\sin\left(\theta_{2}\right),\\
\psi=\tilde{B}_{10}\sin\left(\theta_{1}\right)+\tilde{B}_{01}\sin\left(\theta_{2}\right),\\
\varphi=C_{10}\cos\left(\theta_{1}\right)+C_{01}\cos\left(\theta_{2}\right),
\end{gather}

\noindent where the linear amplitudes satisfy

\begin{alignat}{1}
C_{10} & =\frac{\varepsilon_{1}}{\left(\omega_{1}^{2}-\Delta^{2}k_{1}^{2}\right)\left(1+\frac{\kappa^{2}}{k_{1}^{2}}\right)-1},\\
C_{01} & =\frac{\varepsilon_{2}}{\left(\omega_{2}^{2}-\Delta^{2}k_{2}^{2}\right)\left(1+\frac{\kappa^{2}}{k_{2}^{2}}\right)-1},
\end{alignat}

\begin{alignat}{1}
\tilde{A}_{10} & =-k_{1}\left(1+\frac{\kappa^{2}}{k_{1}^{2}}\right)C_{10},\label{eq:A_lin_1}\\
\tilde{A}_{01} & =-k_{2}\left(1+\frac{\kappa^{2}}{k_{2}^{2}}\right)C_{01},\label{eq:A_lin_2}
\end{alignat}

\begin{alignat}{1}
\tilde{B}_{10} & =-\omega_{1}\left(1+\frac{\kappa^{2}}{k_{1}^{2}}\right)C_{10},\label{eq:B_lin_1}\\
\tilde{B}_{01} & =-\omega_{2}\left(1+\frac{\kappa^{2}}{k_{2}^{2}}\right)C_{01}.\label{eq:B_lin_2}
\end{alignat}

The form of the driving potential together with the linear stage solution
suggest using the following weakly nonlinear ansatz describing the
two-phase solutions for the potentials $\sigma$, $\psi$, and $\varphi$:

\begin{multline}
\sigma=\tilde{A}_{10}\sin\left(\theta_{1}\right)+\tilde{A}_{01}\sin\left(\theta_{2}\right)\\
+\tilde{A}_{11}\sin\left(\theta_{1}+\theta_{2}\right)+\tilde{A}_{1,-1}\sin\left(\theta_{1}-\theta_{2}\right)\\
+\tilde{A}_{20}\sin\left(2\theta_{1}\right)+\tilde{A}_{02}\sin\left(2\theta_{2}\right),\label{eq:sigma_ansatz}
\end{multline}

\begin{multline}
\psi=\tilde{B}_{10}\sin\left(\theta_{1}\right)+\tilde{B}_{01}\sin\left(\theta_{2}\right)\\
+\tilde{B}_{11}\sin\left(\theta_{1}+\theta_{2}\right)+\tilde{B}_{1,-1}\sin\left(\theta_{1}-\theta_{2}\right)\\
+\tilde{B}_{20}\sin\left(2\theta_{1}\right)+\tilde{B}_{02}\sin\left(2\theta_{2}\right),\label{eq:psi_ansatz}
\end{multline}

\begin{multline}
\varphi=C_{10}\cos\left(\theta_{1}\right)+C_{01}\cos\left(\theta_{2}\right)\\
+C_{11}\cos\left(\theta_{1}+\theta_{2}\right)+C_{1,-1}\cos\left(\theta_{1}-\theta_{2}\right)\\
+C_{20}\cos\left(2\theta_{1}\right)+C_{02}\cos\left(2\theta_{2}\right),\label{eq:phi_ansatz}
\end{multline}

\noindent where $\theta_{i}=k_{i}x-\int\omega_{i}\left(t\right)dt$
($i=1,2$) are phases of the solutions.

Note that the above solutions correspond not to the superposition
of two separate nonlinear waves as in Ref.~\citep{Buchanan1993},
but to a single two-phase nonlinear wave, i.e., the solutions have
the form $f(\theta_{1},\theta_{2})$ as opposed to $f_{1}(\theta_{1})+f_{2}(\theta_{2})$.

\noindent 
\begin{figure}
\textbf{(a)}

\includegraphics[width=1\columnwidth]{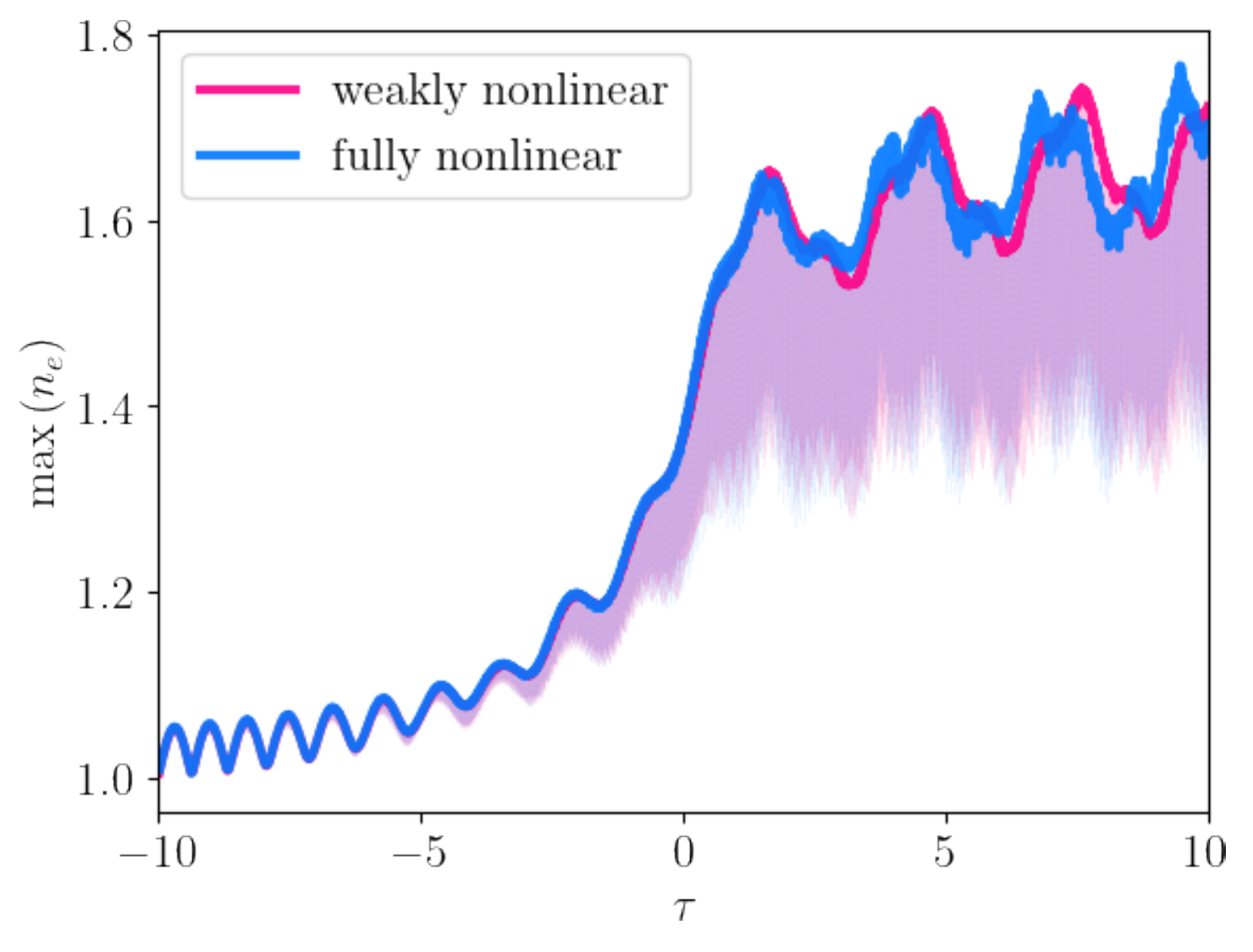}

\textbf{(b)}

\includegraphics[width=1\columnwidth]{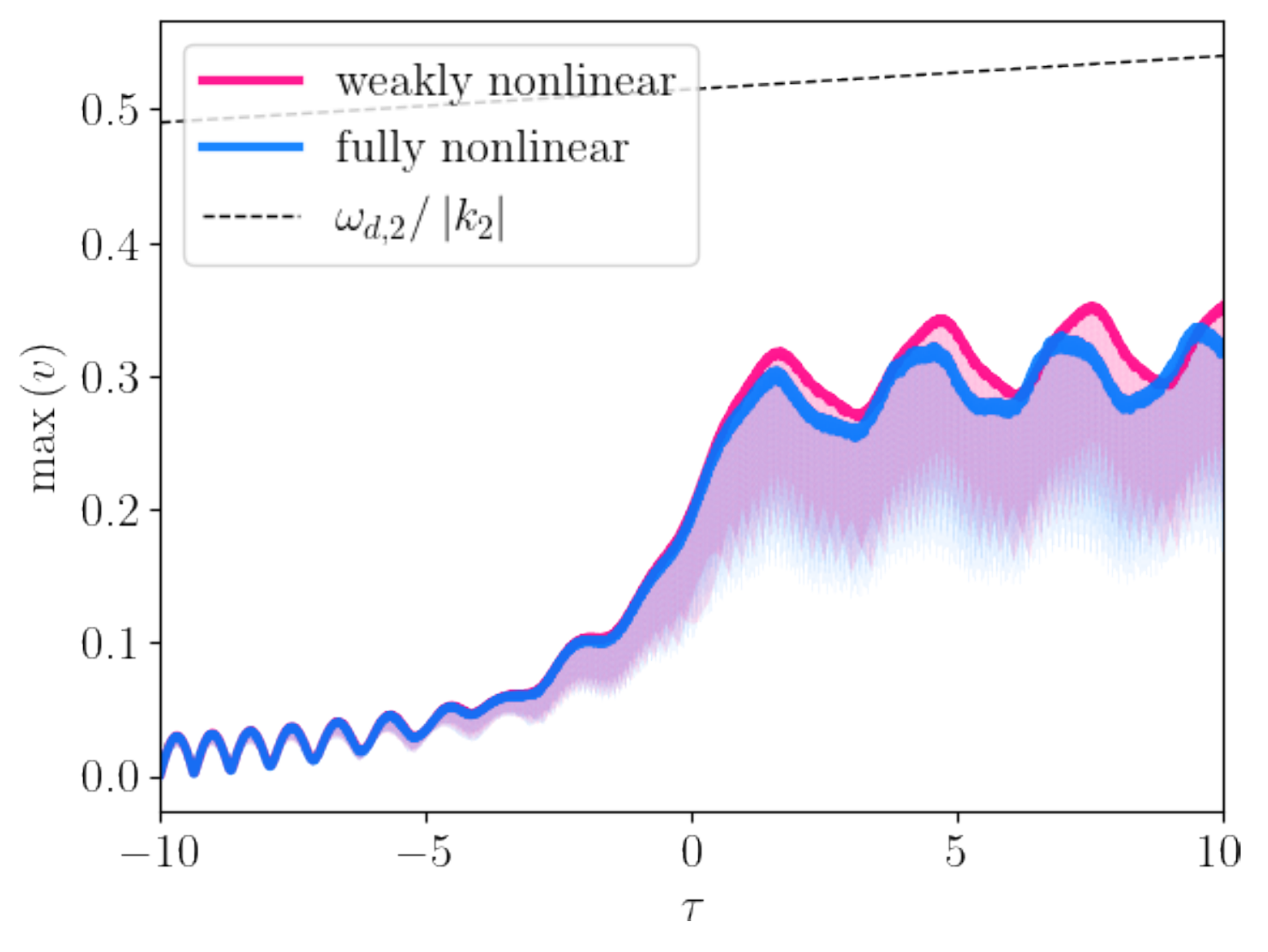}

\caption{\label{fig_max_u_vel}The maximum over $x$ of the electron density
$n_{e}\left(x,\tau\right)$ (a) and the electron fluid velocity $v\left(x,\tau\right)$
(b) vs slow time $\tau=\sqrt{\alpha_{1}}t$ for a two-phase autoresonant
ion acoustic wave excited by two driving counterpropagating traveling
waves with $k_{1}=1$ and $k_{2}=-2$. The solutions were obtained
by solving the fully nonlinear equations~(\ref{eq:sigma_xt})\textendash (\ref{eq:phi_xx_2})
(denoted as \textquotedblleft fully nonlinear\textquotedblright ,
blue line) and the weakly nonlinear equations~(\ref{eq:dI1_dt})\textendash (\ref{eq:dPHI2_dt})
(denoted as \textquotedblleft weakly nonlinear\textquotedblright ,
pink line). The dashed black line represents the absolute value of
the phase velocity of the second driving wave $\omega_{d,2}\left(\tau\right)/\left|k_{2}\right|$
vs $\tau$.}
\end{figure}

To explicitly separate slow and fast phase variables we introduce
phase mismatches $\Phi_{i}=\theta_{i}-\theta_{d,i}$ ($i=1,2$) between
phases of the solutions $\theta_{i}$ and the driving phases $\theta_{d,i}$,
so that the driving term given by Eq.~(\ref{eq:phi_d}) becomes $\varphi_{d}=\varepsilon_{1}\cos\left(\theta_{1}-\Phi_{1}\right)+\varepsilon_{2}\cos\left(\theta_{2}-\Phi_{2}\right)$.
It should be understood then that the coefficients in our ansatz and
phase mismatches $\Phi_{1}$, $\Phi_{2}$ are slow functions of time,
while the phases $\theta_{1}$, $\theta_{2}$ are rapidly varying
functions of time.

According to Whitham's variational principle, we need to obtain the
averaged Lagrangian density $\bar{L}$ by integrating the full Lagrangian
density~(\ref{eq:Lagrangian}) over the rapidly varying phases $\theta_{1}$,
$\theta_{2}$:

\begin{equation}
\bar{L}=\left\langle L\right\rangle _{\theta_{1},\theta_{2}}=\int L\frac{d\theta_{1}}{2\pi}\frac{d\theta_{2}}{2\pi}.
\end{equation}

\noindent The resulting averaged Lagrangian density $\bar{L}$ will
be a function of the slowly varying amplitudes and the phase mismatches
only. This averaged Lagrangian is presented in Appendix~\ref{Appendix_A}.

After obtaining the averaged Lagrangian density $\bar{L}$, we can
use the variational principle $\delta\left(\int\bar{L}dxdt\right)=0$
to derive the weakly nonlinear equations that describe the evolution
of slowly modulated parameters (amplitudes and phase mismatches).

Following Ref.~\citep{Munirov2022a}, we first take variations with
respect to the phases and, after keeping the lowest significant order
terms and using the linear relations (\ref{eq:A_lin_1})\textendash (\ref{eq:B_lin_2}),
we obtain

\begin{align}
\frac{d}{dt}\left[\omega_{1}\left(1+\frac{\kappa^{2}}{k_{1}^{2}}\right)C_{10}^{2}\right]=-\varepsilon_{1}C_{10}\sin\left(\Phi_{1}\right),\label{eq:dC10/dt}\\
\frac{d}{dt}\left[\omega_{2}\left(1+\frac{\kappa^{2}}{k_{2}^{2}}\right)C_{01}^{2}\right]=-\varepsilon_{2}C_{01}\sin\left(\Phi_{2}\right).\label{eq:dC01/dt}
\end{align}

Taking variations of the averaged Lagrangian density $\bar{L}$ with
respect to the first-order amplitudes and expanding around the linear
dispersion relation $\omega_{i}=\omega_{p,i}+\Delta\omega_{i}$ ($i=1,2$),
we obtain

\begin{multline}
\Delta\omega_{1}=-2\omega_{1}k_{1}^{2}\left(1+\frac{\kappa^{2}}{k_{1}^{2}}\right)^{2}a\left(k_{1},\omega_{1}\right)C_{10}^{2}\\
-2\omega_{2}k_{2}^{2}\left(1+\frac{\kappa^{2}}{k_{2}^{2}}\right)^{2}b\left(k_{1},\omega_{1};k_{2},\omega_{2}\right)C_{01}^{2}\\
+\frac{\varepsilon_{1}}{2\omega_{1}\left(1+\frac{\kappa^{2}}{k_{1}^{2}}\right)C_{10}}\cos\left(\Phi_{1}\right),\label{eq:delta_w1}
\end{multline}

\begin{multline}
\Delta\omega_{2}=-2\omega_{2}k_{2}^{2}\left(1+\frac{\kappa^{2}}{k_{2}^{2}}\right)^{2}c\left(k_{2},\omega_{2}\right)C_{01}^{2}\\
-2\omega_{1}k_{1}^{2}\left(1+\frac{\kappa^{2}}{k_{1}^{2}}\right)^{2}b\left(k_{1},\omega_{1};k_{2},\omega_{2}\right)C_{10}^{2}\\
+\frac{\varepsilon_{2}}{2\omega_{2}\left(1+\frac{\kappa^{2}}{k_{2}^{2}}\right)C_{01}}\cos\left(\Phi_{2}\right),\label{eq:delta_w2}
\end{multline}

\noindent where the functions $a\left(k_{1},\omega_{1}\right)$, $b\left(k_{1},\omega_{1};k_{2},\omega_{2}\right)$,
$c\left(k_{2},\omega_{2}\right)$ are defined in Appendix~\ref{Appendix_B}.

Notice that for $\varepsilon_{1}=\varepsilon_{2}=0$, $C_{01}=0$,
and $\kappa=0$, we obtain from Eqs.~(\ref{eq:delta_w1})\textendash (\ref{eq:delta_w2})
the nonlinear frequency shift for a single nonlinear wave in the absence
of the drives:

\begin{equation}
\frac{\Delta\omega_{1}}{\omega_{1}}=k_{1}^{4}\frac{6+9\frac{\Delta^{2}k_{1}^{2}}{\omega_{1}^{2}}+\left(\frac{\Delta^{2}k_{1}^{2}}{\omega_{1}^{2}}\right)^{2}}{12\left(1-\frac{\Delta^{2}k_{1}^{2}}{\omega_{1}^{2}}\right)}C_{10}^{2},
\end{equation}

\noindent which agrees with the nonlinear frequency shift for a single
nonlinear wave in the laboratory frame (see Ref.~\citep{Liu2015}
and references therein).

\noindent 
\begin{figure}[t]
\textbf{(a)}

\includegraphics[width=1\columnwidth]{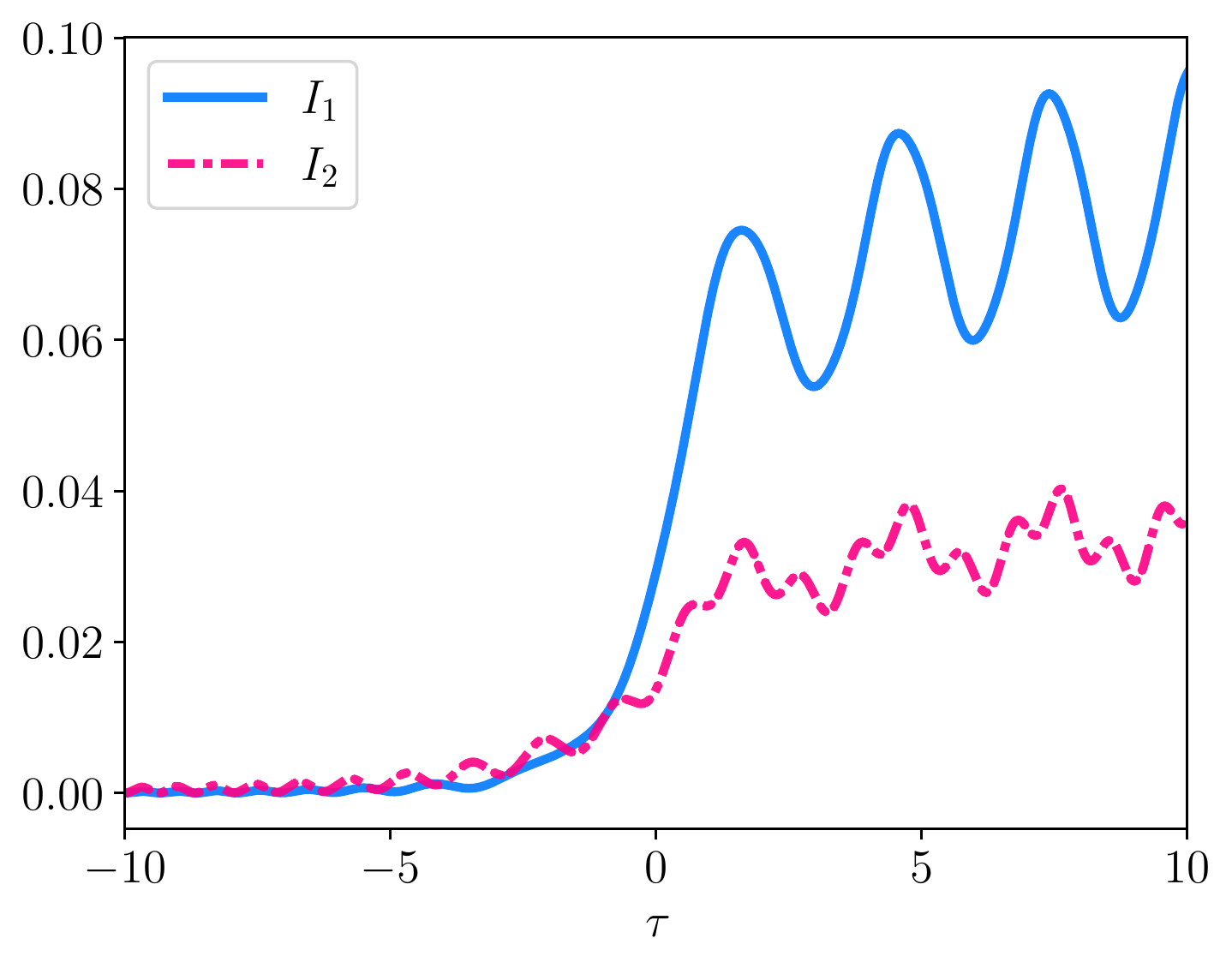}

\textbf{(b)}

\includegraphics[width=1\columnwidth]{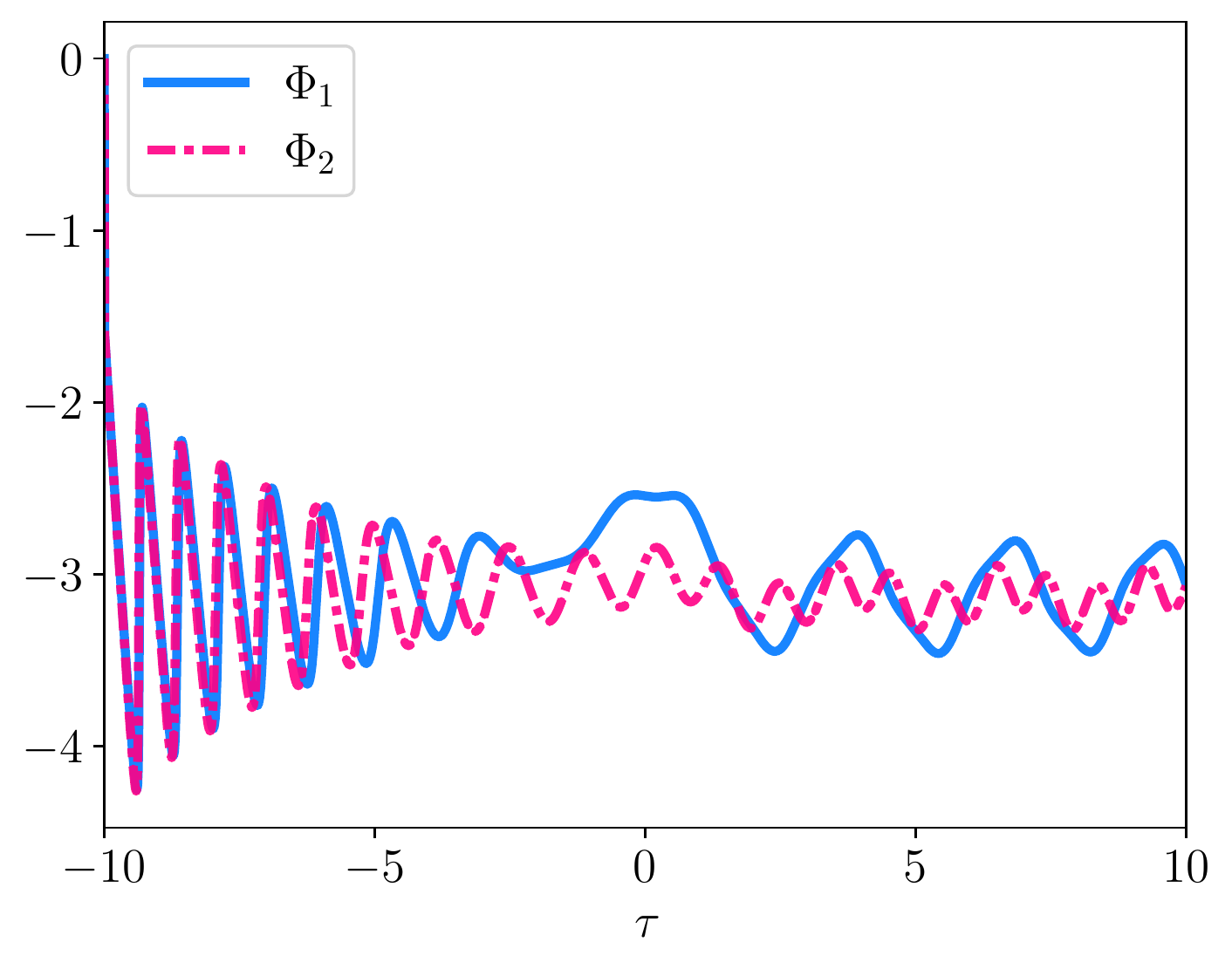}

\caption{\label{fig_I_Phi}The effective actions $I_{1},I_{2}$ (a) and the
phase mismatches $\Phi_{1},\Phi_{2}$ (b) as functions of slow time
$\tau=\sqrt{\alpha_{1}}t$ obtained by solving the weakly nonlinear
equations~(\ref{eq:dI1_dt})\textendash (\ref{eq:dPHI2_dt}).}
\end{figure}

Finally, assuming a slow drive of the form $\omega_{d,i}\left(t\right)=\omega_{p,i}+f_{i}\left(t\right)$
and defining the effective action variables and rescaled amplitudes
through

\begin{gather}
I_{1}=2\omega_{1}k_{1}^{2}\left(1+\frac{\kappa^{2}}{k_{1}^{2}}\right)^{2}C_{10}^{2},\\
I_{2}=2\omega_{2}k_{2}^{2}\left(1+\frac{\kappa^{2}}{k_{2}^{2}}\right)^{2}C_{01}^{2},\\
\epsilon_{1}=-2\left|k_{1}\right|\varepsilon_{1},\;\epsilon_{2}=-2\left|k_{2}\right|\varepsilon_{2},
\end{gather}

\noindent we can rewrite Eqs.~(\ref{eq:dC10/dt})\textendash (\ref{eq:delta_w2})
and obtain the following system of coupled weakly nonlinear evolution
equations:

\begin{alignat}{1}
\frac{dI_{1}}{dt} & =\epsilon_{1}\sqrt{\frac{I_{1}}{2\omega_{1}}}\sin\left(\Phi_{1}\right),\label{eq:dI1_dt}\\
\frac{dI_{2}}{dt} & =\epsilon_{2}\sqrt{\frac{I_{2}}{2\omega_{2}}}\sin\left(\Phi_{2}\right),\label{eq:dI2_dt}\\
\frac{d\Phi_{1}}{dt} & =aI_{1}+bI_{2}+f_{1}\left(t\right)+\frac{\epsilon_{1}}{2\sqrt{2\omega_{1}I_{1}}}\cos\left(\Phi_{1}\right),\label{eq:dPHI1_dt}\\
\frac{d\Phi_{2}}{dt} & =bI_{1}+cI_{2}+f_{2}\left(t\right)+\frac{\epsilon_{2}}{2\sqrt{2\omega_{2}I_{2}}}\cos\left(\Phi_{2}\right).\label{eq:dPHI2_dt}
\end{alignat}

The coupled weakly nonlinear equations~(\ref{eq:dI1_dt})\textendash (\ref{eq:dPHI2_dt})
comprise a system of ordinary differential equations and can be easily
solved using any modern numerical library. Thus, the system of the
weakly nonlinear equations~(\ref{eq:dI1_dt})\textendash (\ref{eq:dPHI2_dt})
allows us to obtain straightforward numerical solutions as well as
to study the conditions for double autoresonance. The possibility
of exciting double autoresonance and limitations on the parameter
space in the systems governed by Eqs.~(\ref{eq:dI1_dt})\textendash (\ref{eq:dPHI2_dt})
were discussed in detail in Ref.~\citep{Munirov2022a} (see also
Ref.~\citep{Barth2007}), so we will not repeat them here. These
results can be easily reproduced for the functional forms of $a\left(k_{1},\omega_{1}\right)$,
$b\left(k_{1},\omega_{1};k_{2},\omega_{2}\right)$, $c\left(k_{2},\omega_{2}\right)$
given in Appendix~\ref{Appendix_B}.

The results of the numerical solution of the weakly nonlinear system~(\ref{eq:dI1_dt})\textendash (\ref{eq:dPHI2_dt})
are presented in Figs.~\ref{fig_n_weakly} and~\ref{fig_vel_weakly}\textendash \ref{fig_I_Phi}.
The parameters used in the simulations are identical to the ones used
in the fully nonlinear numerical simulations of the previous section.
Figure~\ref{fig_n_weakly} shows a colormap of the electron density
$n_{e}\left(x,\tau\right)$ vs slow time $\tau$ and coordinate $x$
(as in Fig.~\ref{fig_n_nl}), while Fig.~\ref{fig_vel_weakly} shows
a similar colormap but for the fluid velocity $v\left(x,\tau\right)$
(as in Fig.~\ref{fig_vel_nl}). After comparing Fig.~\ref{fig_n_nl}
with Fig.~\ref{fig_n_weakly} and Fig.~\ref{fig_vel_nl} with Fig.~\ref{fig_vel_weakly},
we can conclude that the weakly nonlinear theory works well in modeling
the original system. This should be even clearer from Fig.~\ref{fig_max_u_vel},
which compares the maximum values over $x$ of the electron densities
$n_{e}\left(x,\tau\right)$ {[}Fig.~\hyperref[fig_max_u_vel]{\ref{fig_max_u_vel}(a)}{]}
and of the fluid velocities $v\left(x,\tau\right)$ {[}Fig.~\hyperref[fig_max_u_vel]{\ref{fig_max_u_vel}(b)}{]}
in the original fully nonlinear simulations (blue color) and in the
weakly nonlinear model (pink color). Indeed, despite a high degree
of nonlinearity, the agreement is quite decent. Figure~\hyperref[fig_max_u_vel]{\ref{fig_max_u_vel}(b)}
also shows the absolute value of the phase velocity of the second
driving wave with $k_{2}=-2$. We can see that the fluid velocity
is below the absolute values of the phase velocities of the driving
waves, which means we are below the wave breaking limit for the parameters
chosen.

Figure~\ref{fig_I_Phi} shows the effective actions $I_{1},I_{2}$
{[}Fig.~\hyperref[fig_I_Phi]{\ref{fig_I_Phi}(a)}{]} and the phase
mismatches $\Phi_{1},\Phi_{2}$ {[}Fig.~\hyperref[fig_I_Phi]{\ref{fig_I_Phi}(b)}{]}
vs slow time $\tau=\sqrt{\alpha_{1}}t$. We can clearly see that the
phase mismatches oscillate around $-\pi$, signifying phase locking,
while the effective actions $I_{1},I_{2}$ enter the resonance at
$\tau=0$ and then remain in the resonance and grow rapidly.

It is known that the autoresonant phenomenon occurs only when the
driving amplitudes exceed certain threshold values; see Refs.~\citep{Munirov2022a,Barth2007}.
The threshold nature of the autoresonance manifests itself for the
system described in this paper as well. As was discussed in Ref.~\citep{Munirov2022a},
it is difficult to obtain the general analytical result for the double
autoresonance thresholds for the systems described by Eqs.~(\ref{eq:dI1_dt})\textendash (\ref{eq:dPHI2_dt}),
and the thresholds are complicated functions of $\alpha_{1}$, $\alpha_{2}$,
$a$, $b$, $c$. However, the threshold condition for a single-phase
wave, i.e., when one of the driving amplitudes in Eqs.~(\ref{eq:dI1_dt})\textendash (\ref{eq:dPHI2_dt})
vanishes, is well known~\citep{Khasin2003}: 

\begin{equation}
\left|\epsilon_{1}\right|>1.644\sqrt{\frac{\omega_{1}}{2\left|a\left(k_{1},\omega_{1}\right)\right|}}\alpha^{\frac{3}{4}},\label{eq:threshold}
\end{equation}

\noindent where we assumed the presence of the first drive only ($\epsilon_{2}=0$)
and the linear chirp rate $\alpha$.

In the next section, we will use the threshold condition~(\ref{eq:threshold})
to estimate the experimental parameters, such as laser intensity and
laser pulse length required for the autoresonant excitation of plasma
and ion acoustic waves.

\section{\label{sec_IV}Estimates of laser parameters}

In this section we estimate the required laser pulse intensity and
length necessary for autoresonant excitation of large amplitude waves.
We will make estimates of the autoresonant excitation of both plasma
waves discussed in this paper and of ion acoustic waves studied in
Ref.~\citep{Munirov2022a}. Since the threshold conditions for double
autoresonance are difficult to obtain in generality, we will consider
autoresonant excitation of single-phase waves here. Nevertheless,
as indicated by numerical simulations, the single-phase estimates
should be good proxies for multiphase waves as well.

\subsection{Ion acoustic waves}

First, let us consider the case of ion acoustic waves discussed in
Ref.~\citep{Munirov2022a}. In this subsection, all the variables
are as they are defined in Ref.~\citep{Munirov2022a}.

The ponderomotive drive can be created by launching two co- or counterpropagating
laser pulses of similar intensity and duration with varying (chirped)
frequencies; the associated beat wave will produce the ponderomotive
potential with the required properties.

From the single-phase threshold condition {[}Eq.~(\ref{eq:threshold}){]},
using definitions of the dimensionless quantities from Ref.~\citep{Munirov2022a},
we find the following approximate condition for the autoresonance
in the system of ion acoustic waves:

\begin{equation}
\textrm{LHS}\left(I,\lambda,n_{e}/n_{c},T_{e}\right)>\textrm{RHS}\left(k_{1},\alpha\right),\label{eq:LHS_vs_RHS}
\end{equation}

\noindent where we introduced the functions

\begin{flalign}
 & \textrm{LHS}\left(I,\lambda,n_{e}/n_{c},T_{e}\right)=\frac{U_{p}\left[\textrm{eV}\right]}{T_{e}\left[\textrm{eV}\right]},\label{eq:LHS_IAW}\\
 & \textrm{RHS}\left(k_{1},\alpha\right)=1.644\sqrt{\frac{\omega_{1}}{2\left|a\left(k_{1},\omega_{1}\right)\right|}}\frac{\left|k_{1}\right|}{2\left(\omega_{1}^{2}-\Delta^{2}k_{1}^{2}\right)}\alpha^{\frac{3}{4}}.\label{eq:RHS_IAW}
\end{flalign}

Here, $T_{e}$ is the initial electron temperature and $U_{p}$ is
the ponderomotive energy given by

\begin{equation}
U_{p}\left[\textrm{eV}\right]=9.33\times10^{-14}I\left[\frac{\textrm{W}}{\textrm{cm}^{2}}\right]\lambda^{2}\left[\mu\textrm{m}\right]\sqrt{1-\frac{n_{e}}{n_{c}}},\label{eq:U_p}
\end{equation}

\noindent where $I$ is the laser intensity, $\lambda$ is the laser
wavelength, and $n_{e}/n_{c}$ is the ratio of the initial electron
density to the critical plasma density. Inside the function $\textrm{RHS}\left(k_{1},\alpha\right)$
{[}Eq.~(\ref{eq:RHS_IAW}){]} we have the dimensionless wave vector
$k_{1}$ (measured in units of the inverse Debye length $\lambda_{D}^{-1}$)
and the dimensionless frequency $\omega_{1}$ (measured in units of
the ion plasma frequency $\omega_{pi}$) given by the linear ion acoustic
wave dispersion relation (see Ref.~\citep{Munirov2022a}); the definitions
of $a\left(k_{1},\omega_{1}\right)$, $\omega_{1}$, $k_{1}$, $\Delta$,
and $\alpha$ are from Ref.~\citep{Munirov2022a}.

\noindent 
\begin{figure}[t]
\includegraphics[width=1\columnwidth]{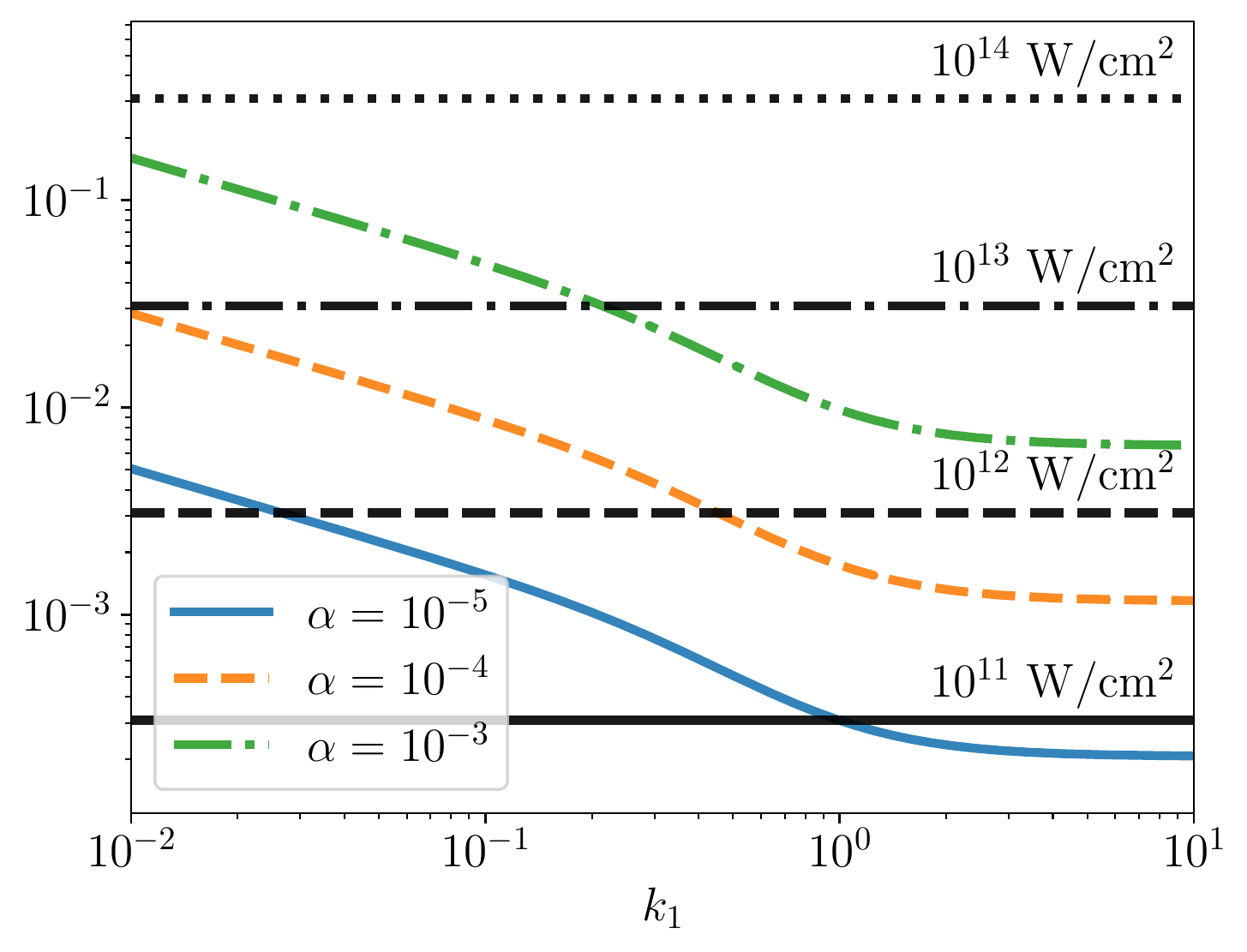}

\caption{\label{fig_threshold_IAW}Approximate thresholds for autoresonant
excitation of single-phase ion acoustic waves. The figure shows the
$\textrm{LHS}\left(I,\lambda=1\:\mu\textrm{m},n_{e}/n_{c}=10^{-2},T_{e}=30\:\textrm{eV}\right)$
given by Eq.~(\ref{eq:LHS_IAW}) for different values of the laser
intensity $I$ {[}$I=10^{11}\:\textrm{W}/\textrm{cm}^{2}$ (solid
black), $I=10^{12}\:\textrm{W}/\textrm{cm}^{2}$ (dashed black), $I=10^{13}\:\textrm{W}/\textrm{cm}^{2}$
(dash-dotted black), and $I=10^{14}\:\textrm{W}/\textrm{cm}^{2}$
(dotted black){]} and the $\textrm{RHS}\left(k_{1},\alpha\right)$
given by Eq.~(\ref{eq:RHS_IAW}) for different values of chirp rate
$\alpha$ {[}$\alpha=10^{-5}$ (solid blue), $\alpha=10^{-4}$ (dashed
orange), and $\alpha=10^{-3}$ (dash-dotted green){]} as functions
of $k_{1}$. For the autoresonance to occur, the RHS must be below
the horizontal line representing the LHS.}
\end{figure}

\noindent 
\begin{figure}[t]
\includegraphics[width=1\columnwidth]{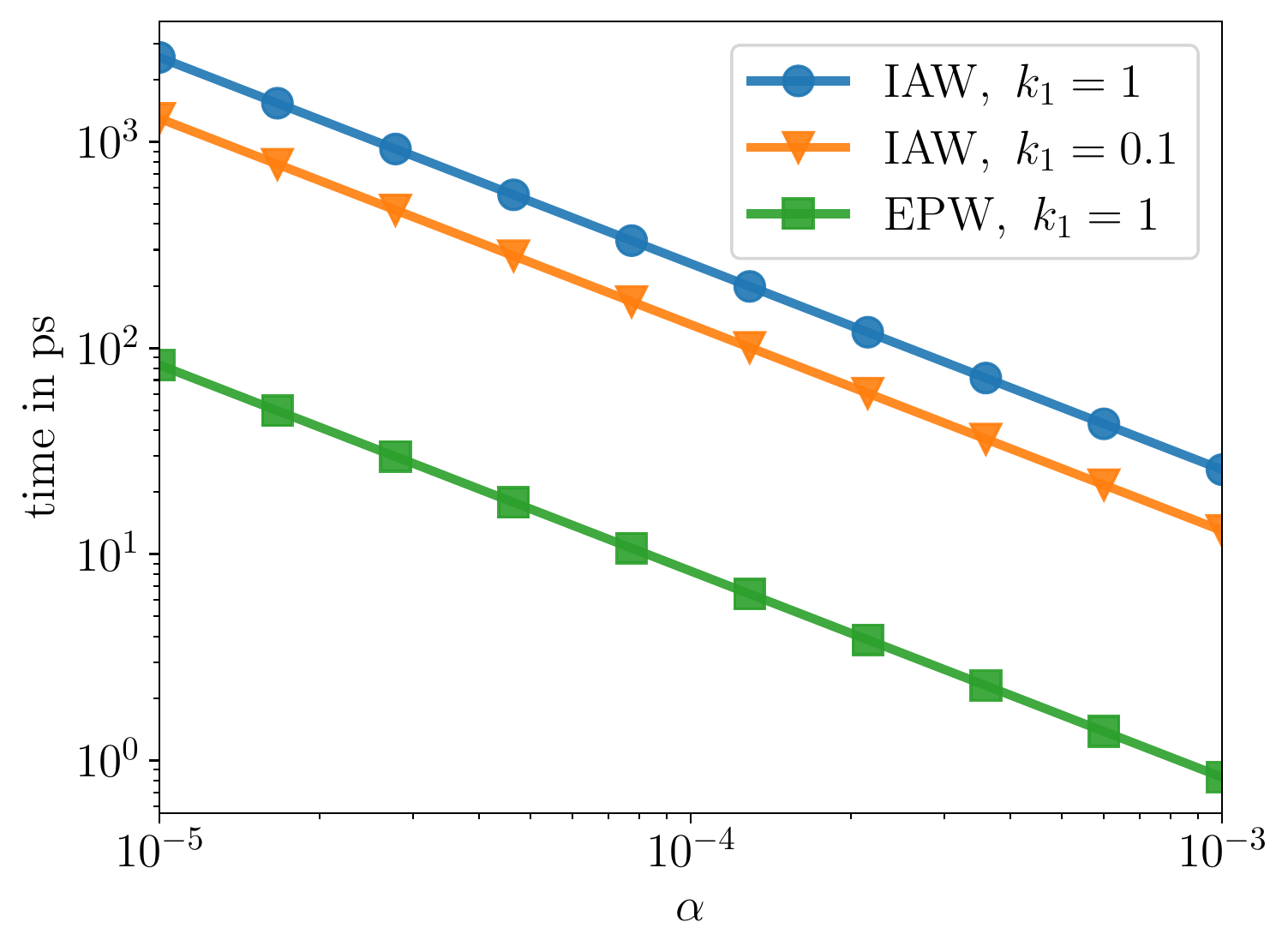}

\caption{\label{fig_threshold_tpulse}Estimate of the laser pulse length required
for the autoresonant excitation of nonlinear waves. The figure shows
$t_{\textrm{pulse}}$ given by Eq.~(\ref{eq:tpulse_IAW}) in picoseconds
required for $\left(\delta n_{e}/n_{e}\right)_{1}$ to reach $0.25$
in autoresonant excitation of ion acoustic waves (IAW) for $k_{1}=0.1$
(orange line with triangle markers) and $k_{1}=1$ (blue line with
circle markers) as functions of chirp rate~$\alpha$. In addition,
the figure shows $t_{\textrm{pulse}}$ given by Eq.~(\ref{eq:tpulse_plasma})
in picoseconds required for $\left(\delta n_{e}/n_{e}\right)_{1}$
to reach $0.25$ in autoresonant excitation of electron plasma waves
(EPW) for $k_{1}=1$ (green line with square markers) as a function
of chirp rate~$\alpha$.}
\end{figure}

\noindent 
\begin{figure}
\includegraphics[width=1\columnwidth]{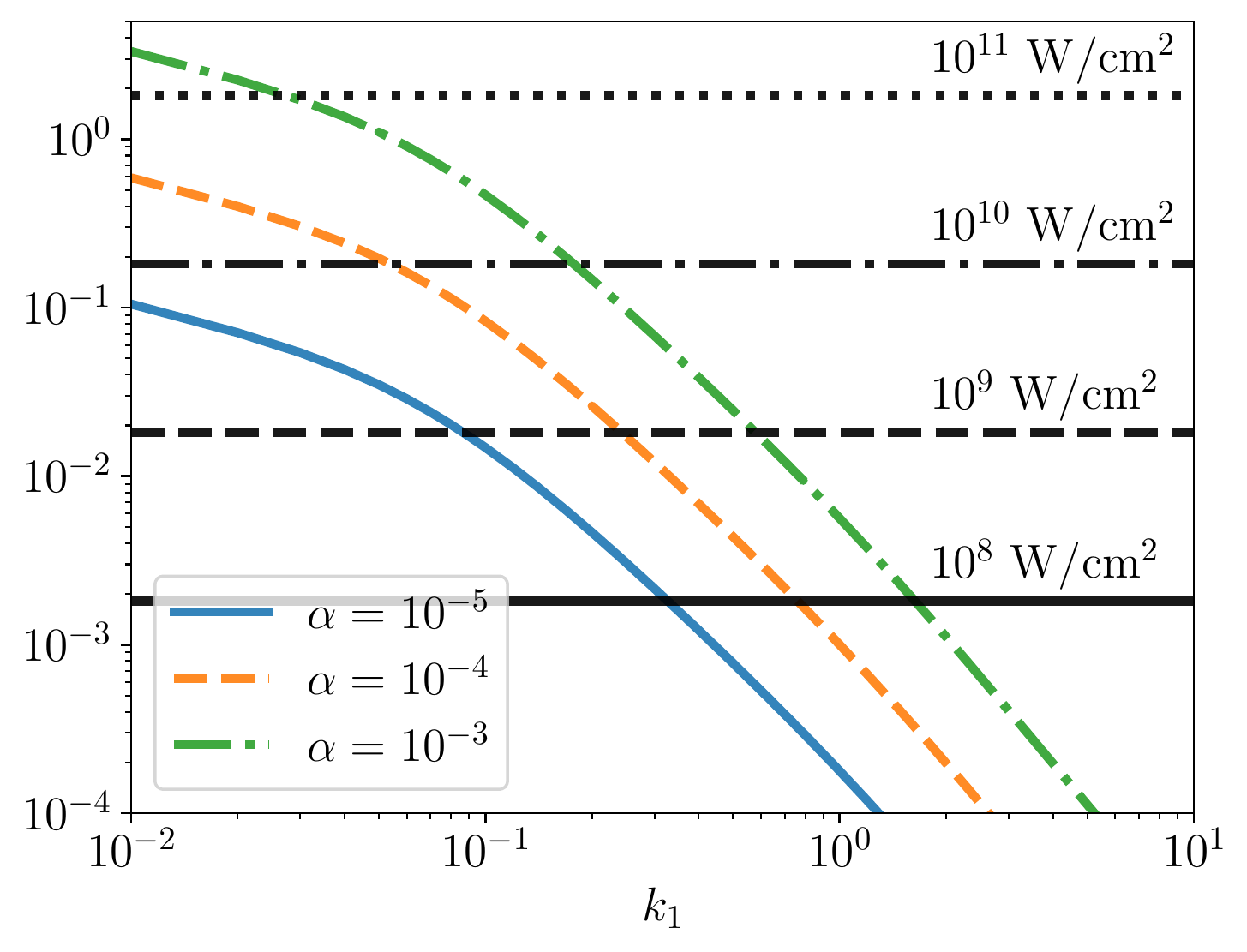}

\caption{\label{fig_threshold_plasma}Approximate thresholds for autoresonant
excitation of single-phase electron plasma waves. The figure shows
the $\textrm{LHS}\left(I,\lambda=1\:\mu\textrm{m},n_{e}/n_{c}=10^{-3},T_{e}=30\:\textrm{eV}\right)$
given by Eq.~(\ref{eq:LHS_plasma}) for different values of the laser
intensity $I$ {[}$I=10^{8}\:\textrm{W}/\textrm{cm}^{2}$ (solid black),
$I=10^{9}\:\textrm{W}/\textrm{cm}^{2}$ (dashed black), $I=10^{10}\:\textrm{W}/\textrm{cm}^{2}$
(dash-dotted black), and $I=10^{11}\:\textrm{W}/\textrm{cm}^{2}$
(dotted black){]} and the $\textrm{RHS}\left(k_{1},\alpha\right)$
given by Eq.~(\ref{eq:RHS_plasma}) for different values of chirp
rate $\alpha$ {[}$\alpha=10^{-5}$ (solid blue), $\alpha=10^{-4}$
(dashed orange), and $\alpha=10^{-3}$ (dash-dotted green){]} as functions
of $k_{1}$. For the autoresonance to occur, the RHS must be below
the horizontal line representing the LHS.}
\end{figure}

Figure~\ref{fig_threshold_IAW} shows the LHS given by Eq.~(\ref{eq:LHS_IAW})
for different values of the laser intensity $I$ and the RHS given
by Eq.~(\ref{eq:RHS_IAW}) for different values of chirp rate $\alpha$
as functions of dimensionless $k_{1}$ (measured in $\lambda_{D}^{-1}$)
for $\lambda=1\:\mu\textrm{m}$, $n_{e}/n_{c}=10^{-2}$, $T_{e}=30\:\textrm{eV}$,
and cold ions ($\Delta=0$). For the autoresonance to occur, the RHS
for given values of $\alpha$ and $k_{1}$ must be below the horizontal
line representing the LHS for a given value of $I$. 

We note that one must take into account additional physical restrictions
on the accessible values of $k_{1}$. For example, to avoid strong
Landau damping, $k_{1}$ cannot be too large. We now estimate the
required laser intensity for two realistic values of the dimensionless
wave vector $k_{1}=0.1$ and $k_{1}=1$. We see from Fig.~\ref{fig_threshold_IAW}
that for the chosen parameters, if $k_{1}=0.1$, autoresonance occurs
when the laser intensity exceeds $I\approx10^{14}\:\textrm{W}/\textrm{cm}^{2}$
for $\alpha=10^{-3}$, $I\approx10^{13}\:\textrm{W}/\textrm{cm}^{2}$
for $\alpha=10^{-4}$, and $I\approx10^{12}\:\textrm{W}/\textrm{cm}^{2}$
for $\alpha=10^{-5}$, while if $k_{1}=1$, the intensity should exceed
$I\approx10^{13}\:\textrm{W}/\textrm{cm}^{2}$ for $\alpha=10^{-3}$,
$I\approx10^{12}\:\textrm{W}/\textrm{cm}^{2}$ for $\alpha=10^{-4}$,
and $I\approx10^{11}\:\textrm{W}/\textrm{cm}^{2}$ for $\alpha=10^{-5}$.

Now let us estimate the laser duration required to autoresonantly
excite large amplitude waves. Since the electron density in Ref.~\citep{Munirov2022a}
can be approximated as $e^{\varphi}$, we can estimate in the leading
linear order the relative density increase as $\left(\delta n_{e}\right)_{1}\thickapprox\left(\delta\varphi\right)_{1}\thickapprox C_{10}$.
Then, using the asymptotic solution for the effective action $\bar{I}_{1}\approx\left[\alpha/\left|a\left(k_{1},\omega_{1}\right)\right|\right]t$
and its connection with $C_{10}$ (see definitions of Ref.~\citep{Munirov2022a}),
we can estimate the dimensionless laser pulse length (measured in
units of $\omega_{pi}^{-1}$) required to reach $\left(\delta n_{e}/n_{e}\right)_{1}$
as

\begin{equation}
t_{\textrm{pulse}}=\left[\left(\frac{\delta n_{e}}{n_{e}}\right)_{1}\right]^{2}\frac{2\omega_{1}k_{1}^{2}}{\left(\omega_{1}^{2}-\Delta^{2}k_{1}^{2}\right)^{2}}\frac{\left|a\left(k_{1},\omega_{1}\right)\right|}{\alpha},\label{eq:tpulse_IAW}
\end{equation}

\noindent where the definitions of $a\left(k_{1},\omega_{1}\right)$,
$\omega_{1}$, $k_{1}$, $\Delta$, and $\alpha$ are from Ref.~\citep{Munirov2022a}.

Figure~\ref{fig_threshold_tpulse} plots the pulse length determined
by Eq.~(\ref{eq:tpulse_IAW}) in picoseconds as a function of chirp
rate $\alpha$ for $\left(\delta n_{e}/n_{e}\right)_{1}=0.25$ for
$k_{1}=0.1$ (orange line with triangle markers) and $k_{1}=1$ (blue
line with circle markers). Other parameters are the same as in Fig.~\ref{fig_threshold_IAW},
namely, $\lambda=1\:\mu\textrm{m}$, $n_{e}/n_{c}=10^{-2}$, $T_{e}=30\:\textrm{eV}$,
$\Delta=0$. We can see that for $\alpha=10^{-4}$ the laser pulse
length of $t_{\textrm{pulse}}\approx\textrm{10--100}\:\textrm{ps}$
is required, while for $\alpha=10^{-3}$ the laser pulse length of
$t_{\textrm{pulse}}\approx\textrm{1--10}\:\textrm{ps}$ should be
sufficient.

We note that the chosen linear increase in the relative density $\left(\delta n_{e}/n_{e}\right)_{1}=0.25$
corresponds in practice to large density fluctuations on the order
of $\delta n_{e}/n_{e}\sim1$. We also note that even though for the
two-phase case there are additional restrictions on the values of
$\varepsilon_{1}$, $\varepsilon_{2}$, $\alpha_{1}$, $\alpha_{2}$,
$k_{1}$, and $k_{2}$, the actual requirements for the laser intensity
and duration can be even lower than for the single-phase case, because
the excited electron density in the ``interference'' pattern of
the two drives can be larger than for the individual drives. Indeed,
this can be seen by comparing $\delta n_{e}/n_{e}\approx0.2$ {[}Figs.~\hyperref[fig_n_nl]{\ref{fig_n_nl}(b)}
and~\hyperref[fig_n_nl]{\ref{fig_n_nl}(c)}{]} for a single-phase
excitation with $\delta n_{e}/n_{e}\approx0.8$ {[}Fig.~\hyperref[fig_n_nl]{\ref{fig_n_nl}(a)}{]}
for a two-phase excitation.

Thus, we can expect that the laser intensities on the order of $I\approx10^{12}\textrm{--}10^{14}\:\textrm{W}/\textrm{cm}^{2}$
and the laser duration on the order of $t_{\textrm{pulse}}\approx\textrm{1--1000}\:\textrm{ps}$
should be sufficient to autoresonantly drive a single-phase ion acoustic
wave.

\subsection{Electron plasma waves}

Now we can make similar estimates but for the case of electron plasma
waves discussed in this paper. In this subsection, all the variables
are as they are defined in the current paper.

The single-phase threshold condition~(\ref{eq:threshold}) can again
be presented in the form given by Eq.~(\ref{eq:LHS_vs_RHS}), but
with the following definitions for the functions $\textrm{LHS}$ and
$\textrm{RHS}$:

\begin{flalign}
 & \textrm{LHS}\left(I,\lambda,n_{e}/n_{c},T_{e}\right)=\frac{U_{p}\left[\textrm{eV}\right]}{m_{e}\omega_{p}^{2}/k^{2}\left[\textrm{eV}\right]},\label{eq:LHS_plasma}\\
 & \textrm{RHS}\left(k_{1},\alpha\right)=1.644\sqrt{\frac{\omega_{1}}{2\left|a\left(k_{1},\omega_{1}\right)\right|}}\frac{1}{2\left|k_{1}\right|}\alpha^{\frac{3}{4}}.\label{eq:RHS_plasma}
\end{flalign}

Here, $U_{p}$ is the ponderomotive energy given by Eq.~(\ref{eq:U_p}).
Inside the function $\textrm{RHS}\left(k_{1},\alpha\right)$ {[}Eq.~(\ref{eq:RHS_plasma}){]}
we have the dimensionless wave vector $k_{1}$ (measured in $k$)
and the dimensionless frequency $\omega_{1}$ (measured in units of
the electron plasma frequency $\omega_{p}$) given by the linear plasma
wave dispersion relation {[}Eq.~(\ref{eq:w_p}){]}; the definition
of the function $a\left(k_{1},\omega_{1}\right)$ is from Appendix~\ref{Appendix_B},
while $\omega_{1}$, $k_{1}$, $\kappa$, $\Delta$, and $\alpha$
are as they are defined in this paper.

Figure~\ref{fig_threshold_plasma} shows the LHS given by Eq.~(\ref{eq:LHS_plasma})
for different values of the laser intensity $I$ and the RHS given
by Eq.~(\ref{eq:RHS_plasma}) for different values of chirp rate
$\alpha$ as functions of the dimensionless $k_{1}$ (measured in
units of $k$) for $\lambda=1\:\mu\textrm{m}$, $n_{e}/n_{c}=10^{-3}$,
$T_{e}=30\:\textrm{eV}$, $\kappa=0.1$. Since we consider in this
estimate the threshold for one wave, we measure $k_{1}$ in the wave
vector of the drive; thus, the relevant value of $k_{1}$ is $k_{1}=1$.
We see from Fig.~\ref{fig_threshold_plasma} that autoresonance should
occur if the laser intensity exceeds $I\approx10^{10}\:\textrm{W}/\textrm{cm}^{2}$
for $\alpha=10^{-3}$, $I\approx10^{9}\:\textrm{W}/\textrm{cm}^{2}$
for $\alpha=10^{-4}$, and $I\approx10^{8}\:\textrm{W}/\textrm{cm}^{2}$
for $\alpha=10^{-5}$.

As in the case of ion acoustic waves, we can estimate the required
laser pulse duration for the autoresonant excitation of electron plasma
waves. In the leading linear order, the relative density increase
is $\left(\delta n_{e}\right)_{1}\thickapprox k_{1}\widetilde{A}_{10}$.
Then, using the asymptotic solution for the effective action $\bar{I}_{1}\approx\left[\alpha/\left|a\left(k_{1},\omega_{1}\right)\right|\right]t$
and its connection with $\widetilde{A}_{10}$, we can estimate the
dimensionless laser pulse length (measured in units of $\omega_{p}^{-1}$)
required to reach $\left(\delta n_{e}/n_{e}\right)_{1}$ as

\begin{equation}
t_{\textrm{pulse}}=\left[\left(\frac{\delta n_{e}}{n_{e}}\right)_{1}\right]^{2}\frac{2\omega_{1}\left|a\left(k_{1},\omega_{1}\right)\right|}{\alpha k_{1}^{2}},\label{eq:tpulse_plasma}
\end{equation}

\noindent where the definition of the function $a\left(k_{1},\omega_{1}\right)$
is from Appendix~\ref{Appendix_B}, while $\omega_{1}$, $k_{1}$,
$\kappa$, $\Delta$, and $\alpha$ are as they are defined in this
paper.

Figure~\ref{fig_threshold_tpulse} plots the pulse length determined
by Eq.~(\ref{eq:tpulse_plasma}) in picoseconds as a function of
chirp rate $\alpha$ for $\left(\delta n_{e}/n_{e}\right)_{1}=0.25$
(green line with square markers). Other parameters are the same as
in Fig.~\ref{fig_threshold_plasma}, namely, $\lambda=1\:\mu\textrm{m}$,
$n_{e}/n_{c}=10^{-3}$, $T_{e}=30\:\textrm{eV}$, $\kappa=0.1$. We
can see from Fig.~\ref{fig_threshold_tpulse} that for $\alpha=10^{-5}$
the laser pulse length of $t_{\textrm{pulse}}\approx100\:\textrm{ps}$
is required, for $\alpha=10^{-4}$ the pulse length is estimated as
$t_{\textrm{pulse}}\approx10\:\textrm{ps}$, and for $\alpha=10^{-3}$
the pulse length of $t_{\textrm{pulse}}\approx1\:\textrm{ps}$ should
be sufficient.

Thus, we can expect that the laser intensities on the order of $I\approx10^{9}\textrm{--}10^{11}\:\textrm{W}/\textrm{cm}^{2}$
and the laser duration on the order of $t_{\textrm{pulse}}\approx\textrm{1--100}\:\textrm{ps}$
should be sufficient to autoresonantly drive a single-phase electron
plasma wave.

\section{\label{sec_Conclusions}Conclusions}

We have shown how to use phase locking (autoresonance) with small
amplitude chirped-frequency ponderomotive drives to create and control
strongly nonlinear two-phase plasma waves. The drives can be controlled
independently as long as the conditions for the double autoresonance
are met. We have illustrated these nonlinear two-phase waves through
fully nonlinear numerical simulations. Using Whitham's averaged Lagrangian
procedure we analytically developed a reduced set of ordinary differential
equations for the amplitudes and phases of the waves. This analytical
weakly nonlinear theory is necessary to understand how to choose the
appropriate parameters to drive and control such two-phase structures.

Similar to the case of ion acoustic waves \citep{Munirov2022a}, the
autoresonantly excited multiphase waves form coherent spatiotemporal
quasicrystalline structures, whose properties as accelerating structures
and optical elements require further investigation. These nonlinear
two-phase structures, have not been seen, to our knowledge, in experiments.
The autoresonant excitation described here requires a balance between
the pulse amplitude and chirp rate, as given by the threshold, and
is also constrained by physics not in our model.

We have made initial estimates for the required laser intensities
and pulse lengths. These estimates suggest that the autoresonant method
of creating large amplitude coherent structures in plasmas is promising
but requires additional investigation. Should the autoresonant method
of exciting plasma structures prove to be effective, it would allow
for large amplitude structures to be excited with relatively low intensity
and energy lasers. First principles models such as particle-in-cell
(PIC) simulations will be necessary to further establish the practical
aspects of the experimental realization of the autoresonant electron
plasma or ion acoustic waves, to gauge the influence of other possible
effects (collisional and collisionless damping, various instabilities,
higher dimensionality effects, kinetic effects, such as trapping,
etc.), and to study their long-term stability.

\section*{Acknowledgments}

This work was supported by NSF-BSF Grant No. 1803874 and U.S.-Israel
Binational Science Foundation Grant No. 2020233.

\begin{widetext}

\appendix

\section{The averaged Lagrangian density}
\label{Appendix_A}

The averaged Lagrangian density $\bar{L}$ is the sum of the following
terms:

\begin{equation}
\left\langle \frac{1}{2}\varphi_{x}^{2}\right\rangle _{\theta_{1},\theta_{2}}=\frac{1}{4}k_{1}^{2}C_{10}^{2}+\frac{1}{4}k_{2}^{2}C_{01}^{2}+\frac{1}{4}\left(k_{1}+k_{2}\right)^{2}C_{11}^{2}+\frac{1}{4}\left(k_{1}-k_{2}\right)^{2}C_{1,-1}^{2}+k_{1}^{2}C_{20}^{2}+k_{2}^{2}C_{02}^{2},
\end{equation}

\begin{equation}
\left\langle \frac{1}{2}\kappa^{2}\varphi^{2}\right\rangle _{\theta_{1},\theta_{2}}=\frac{1}{4}\kappa^{2}C_{10}^{2}+\frac{1}{4}\kappa^{2}C_{01}^{2}+\frac{1}{4}\kappa^{2}C_{11}^{2}+\frac{1}{4}\kappa^{2}C_{1,-1}^{2}+\frac{1}{4}\kappa^{2}C_{20}^{2}+\frac{1}{4}\kappa^{2}C_{02}^{2},
\end{equation}

\begin{multline}
\left\langle -\frac{1}{2}\left(\psi_{t}\sigma_{x}+\psi_{x}\sigma_{t}\right)\right\rangle _{\theta_{1},\theta_{2}}=\frac{1}{2}\omega_{1}k_{1}\tilde{B}_{10}\tilde{A}_{10}+\frac{1}{2}\omega_{2}k_{2}\tilde{B}_{01}\tilde{A}_{01}\\
+2\omega_{1}k_{1}\tilde{B}_{20}\tilde{A}_{20}+2\omega_{2}k_{2}\tilde{B}_{02}\tilde{A}_{02}\\
+\frac{1}{2}\left(\omega_{1}-\omega_{2}\right)\left(k_{1}-k_{2}\right)\tilde{B}_{1,-1}\tilde{A}_{1,-1}+\frac{1}{2}\left(\omega_{1}+\omega_{2}\right)\left(k_{1}+k_{2}\right)\tilde{B}_{11}\tilde{A}_{11},
\end{multline}

\begin{multline}
\left\langle -\frac{1}{2}\psi_{x}^{2}\left(1+\sigma_{x}\right)\right\rangle _{\theta_{1},\theta_{2}}=-\frac{1}{4}k_{1}^{2}\tilde{B}_{10}^{2}-\frac{1}{4}k_{2}^{2}\tilde{B}_{01}^{2}-k_{1}^{2}\tilde{B}_{20}^{2}-k_{2}^{2}\tilde{B}_{02}^{2}-\frac{1}{4}\left(k_{1}+k_{2}\right)^{2}\tilde{B}_{11}^{2}-\frac{1}{4}\left(k_{1}-k_{2}\right)^{2}\tilde{B}_{1,-1}^{2}\\
-\frac{1}{2}k_{2}^{3}\left(\tilde{A}_{01}\tilde{B}_{01}\tilde{B}_{02}+\frac{1}{2}\tilde{A}_{02}\tilde{B}_{01}^{2}\right)-\frac{1}{2}k_{1}^{3}\left(\tilde{A}_{10}\tilde{B}_{10}\tilde{B}_{20}+\frac{1}{2}\tilde{A}_{20}\tilde{B}_{10}^{2}\right)\\
-\frac{1}{4}k_{1}k_{2}\left(k_{1}-k_{2}\right)\left(\tilde{A}_{01}\tilde{B}_{10}\tilde{B}_{1,-1}+\tilde{A}_{1,-1}\tilde{B}_{01}\tilde{B}_{10}+\tilde{A}_{10}\tilde{B}_{01}\tilde{B}_{1,-1}\right)\\
-\frac{1}{4}k_{1}k_{2}\left(k_{1}+k_{2}\right)\left(\tilde{A}_{01}\tilde{B}_{11}\tilde{B}_{10}+\tilde{A}_{10}\tilde{B}_{01}\tilde{B}_{11}+\tilde{A}_{11}\tilde{B}_{01}\tilde{B}_{10}\right),
\end{multline}

\begin{multline}
\left\langle -\frac{1}{2}\Delta^{2}\sigma_{x}^{2}\left(1+\frac{1}{3}\sigma_{x}\right)\right\rangle _{\theta_{1},\theta_{2}}=-\frac{1}{4}\Delta^{2}k_{1}^{2}\tilde{A}_{10}^{2}-\frac{1}{4}\Delta^{2}k_{1}^{3}\tilde{A}_{10}^{2}\tilde{A}_{20}-\Delta^{2}k_{1}^{2}\tilde{A}_{20}^{2}-\frac{1}{4}\Delta^{2}k_{2}^{2}\tilde{A}_{01}^{2}-\frac{1}{4}\Delta^{2}k_{2}^{3}\tilde{A}_{01}^{2}\tilde{A}_{02}-\Delta^{2}k_{2}^{2}\tilde{A}_{02}^{2}\\
-\frac{1}{4}\Delta^{2}k_{1}k_{2}\left(k_{1}-k_{2}\right)\tilde{A}_{01}\tilde{A}_{10}\tilde{A}_{1,-1}-\frac{1}{4}\Delta^{2}k_{1}k_{2}\left(k_{1}+k_{2}\right)\tilde{A}_{01}\tilde{A}_{10}\tilde{A}_{11}-\frac{1}{4}\Delta^{2}\left(k_{1}-k_{2}\right)^{2}\tilde{A}_{1,-1}^{2}-\frac{1}{4}\Delta^{2}\left(k_{1}+k_{2}\right)^{2}\tilde{A}_{11}^{2},
\end{multline}

\begin{equation}
\left\langle \sigma_{x}\varphi\right\rangle _{\theta_{1},\theta_{2}}=\frac{1}{2}k_{1}\tilde{A}_{10}C_{10}+\frac{1}{2}k_{2}\tilde{A}_{01}C_{01}+\frac{1}{2}\left(k_{1}+k_{2}\right)\tilde{A}_{11}C_{11}+\frac{1}{2}\left(k_{1}-k_{2}\right)\tilde{A}_{1,-1}C_{1,-1}+k_{1}\tilde{A}_{20}C_{20}+k_{2}\tilde{A}_{02}C_{02},
\end{equation}

\begin{equation}
\left\langle \sigma_{x}\varphi_{d}\right\rangle _{\theta_{1},\theta_{2}}=\frac{1}{2}\varepsilon_{1}k_{1}\tilde{A}_{10}\cos\left(\Phi_{1}\right)+\frac{1}{2}\varepsilon_{2}k_{2}\tilde{A}_{01}\cos\left(\Phi_{2}\right).
\end{equation}

\section{Functions $a\left(k_{1},\omega_{1}\right)$, $b\left(k_{1},\omega_{1};k_{2},\omega_{2}\right)$, and $c\left(k_{2},\omega_{2}\right)$}
\label{Appendix_B}

The function $a\left(k_{1},\omega_{1}\right)$ is defined via

\begin{equation}
C_{10}^{2}a\left(k_{1},\omega_{1}\right)=-\frac{1}{2\omega_{1}\left(1+\frac{\kappa^{2}}{k_{1}^{2}}\right)^{2}}\tilde{B}_{20}-\frac{\omega_{1}^{2}+\Delta^{2}k_{1}^{2}}{4\omega_{1}^{2}k_{1}\left(1+\frac{\kappa^{2}}{k_{1}^{2}}\right)^{2}}\tilde{A}_{20},
\end{equation}

The function $c\left(k_{2},\omega_{2}\right)$ is defined via

\begin{equation}
C_{01}^{2}c\left(k_{2},\omega_{2}\right)=-\frac{1}{2\omega_{2}\left(1+\frac{\kappa^{2}}{k_{2}^{2}}\right)^{2}}\tilde{B}_{02}-\frac{\omega_{2}^{2}+\Delta^{2}k_{2}^{2}}{4\omega_{2}^{2}k_{2}\left(1+\frac{\kappa^{2}}{k_{2}^{2}}\right)^{2}}\tilde{A}_{02},
\end{equation}

\noindent where we note that, due to symmetry, the functions $a\left(k_{1},\omega_{1}\right)$
and $c\left(k_{2},\omega_{2}\right)$ should have an identical functional
dependence.

The function $b\left(k_{1},\omega_{1};k_{2},\omega_{2}\right)$ is
defined via

\begin{multline}
C_{10}C_{01}b\left(k_{1},\omega_{1};k_{2},\omega_{2}\right)=-\frac{\omega_{1}\omega_{2}+\Delta^{2}k_{1}k_{2}}{8\omega_{1}\omega_{2}k_{1}k_{2}\left(1+\frac{\kappa^{2}}{k_{1}^{2}}\right)\left(1+\frac{\kappa^{2}}{k_{2}^{2}}\right)}\left[\left(k_{1}-k_{2}\right)\tilde{A}_{1,-1}+\left(k_{1}+k_{2}\right)\tilde{A}_{11}\right]\\
-\frac{\omega_{1}k_{2}+\omega_{2}k_{1}}{8\omega_{1}\omega_{2}k_{1}k_{2}\left(1+\frac{\kappa^{2}}{k_{1}^{2}}\right)\left(1+\frac{\kappa^{2}}{k_{2}^{2}}\right)}\left[\left(k_{1}-k_{2}\right)\tilde{B}_{1,-1}+\left(k_{1}+k_{2}\right)\tilde{B}_{11}\right].
\end{multline}

Here, the amplitudes $\tilde{A}_{20}$, $\tilde{A}_{02}$, $\tilde{B}_{20}$,
$\tilde{B}_{02}$, $C_{20}$, $C_{02}$, $\tilde{A}_{11}$, $\tilde{A}_{1,-1}$,
$\tilde{B}_{11}$, $\tilde{B}_{1,-1}$, $C_{11}$, $C_{1,-1}$ should
be expressed through $C_{10}$, $C_{01}$, $k_{1}$, $k_{2}$, $\omega_{1}$,
$\omega_{2}$, $\kappa$, $\Delta$ using Eqs.~(\ref{eq:A20_final})\textendash (\ref{eq:C1-1_final})
of Appendix~\ref{Appendix_C}, so that $a\left(k_{1},\omega_{1}\right)$,
$b\left(k_{1},\omega_{1};k_{2},\omega_{2}\right)$, and $c\left(k_{2},\omega_{2}\right)$
are functions of $k_{1}$, $k_{2}$, $\omega_{1}$, $\omega_{2}$,
$\kappa$, $\Delta$ only. The dimensionless frequencies $\omega_{1}$
and $\omega_{2}$ are determined by the linear plasma wave dispersion
relation {[}Eq.~(\ref{eq:w_p}){]}.

\section{The second-order amplitudes}
\label{Appendix_C}

To express the second-order amplitudes through the first-order amplitudes
$C_{10}$ and $C_{01}$, we calculate the variations of the averaged
Lagrangian density $\bar{L}$ with respect to the second-order amplitudes
and, after solving the resulting system of equations and using the
linear relations (\ref{eq:A_lin_1})\textendash (\ref{eq:B_lin_2}),
we obtain the expressions for the second-order amplitudes.

From variations with respect to $\tilde{A}_{20}$, $\tilde{A}_{02}$,
$\tilde{B}_{20}$, $\tilde{B}_{02}$, $C_{20}$, $C_{02}$ and the
linear relations (\ref{eq:A_lin_1})\textendash (\ref{eq:B_lin_2}),
we obtain

\begin{alignat}{1}
\tilde{A}_{20} & =-k_{1}^{3}\left(4+\frac{\kappa^{2}}{k_{1}^{2}}\right)\left(1+\frac{\kappa^{2}}{k_{1}^{2}}\right)^{2}\frac{3\omega_{1}^{2}+\Delta^{2}k_{1}^{2}}{8\left[1-\left(\omega_{1}^{2}-\Delta^{2}k_{1}^{2}\right)\left(4+\frac{\kappa^{2}}{k_{1}^{2}}\right)\right]}C_{10}^{2},\label{eq:A20_final}\\
\tilde{A}_{02} & =-k_{2}^{3}\left(4+\frac{\kappa^{2}}{k_{2}^{2}}\right)\left(1+\frac{\kappa^{2}}{k_{2}^{2}}\right)^{2}\frac{3\omega_{2}^{2}+\Delta^{2}k_{2}^{2}}{8\left[1-\left(\omega_{2}^{2}-\Delta^{2}k_{2}^{2}\right)\left(4+\frac{\kappa^{2}}{k_{2}^{2}}\right)\right]}C_{01}^{2},\label{eq:A02_final}
\end{alignat}

\begin{alignat}{1}
\tilde{B}_{20} & =-\omega_{1}k_{1}^{2}\left(1+\frac{\kappa^{2}}{k_{1}^{2}}\right)^{2}\frac{2+\left(\omega_{1}^{2}+3\Delta^{2}k_{1}^{2}\right)\left(4+\frac{\kappa^{2}}{k_{1}^{2}}\right)}{8\left[1-\left(\omega_{1}^{2}-\Delta^{2}k_{1}^{2}\right)\left(4+\frac{\kappa^{2}}{k_{1}^{2}}\right)\right]}C_{10}^{2},\label{eq:B20_final}\\
\tilde{B}_{02} & =-\omega_{2}k_{2}^{2}\left(1+\frac{\kappa^{2}}{k_{2}^{2}}\right)^{2}\frac{2+\left(\omega_{2}^{2}+3\Delta^{2}k_{2}^{2}\right)\left(4+\frac{\kappa^{2}}{k_{2}^{2}}\right)}{8\left[1-\left(\omega_{2}^{2}-\Delta^{2}k_{2}^{2}\right)\left(4+\frac{\kappa^{2}}{k_{2}^{2}}\right)\right]}C_{01}^{2},\label{eq:B02_final}
\end{alignat}

\begin{alignat}{1}
C_{20} & =k_{1}^{2}\left(1+\frac{\kappa^{2}}{k_{1}^{2}}\right)^{2}\frac{3\omega_{1}^{2}+\Delta^{2}k_{1}^{2}}{4\left[1-\left(\omega_{1}^{2}-\Delta^{2}k_{1}^{2}\right)\left(4+\frac{\kappa^{2}}{k_{1}^{2}}\right)\right]}C_{10}^{2},\label{eq:C20_final}\\
C_{02} & =k_{2}^{2}\left(1+\frac{\kappa^{2}}{k_{2}^{2}}\right)^{2}\frac{3\omega_{2}^{2}+\Delta^{2}k_{2}^{2}}{4\left[1-\left(\omega_{2}^{2}-\Delta^{2}k_{2}^{2}\right)\left(4+\frac{\kappa^{2}}{k_{2}^{2}}\right)\right]}C_{01}^{2}.\label{eq:C02_final}
\end{alignat}

From variations with respect to $\tilde{A}_{11}$, $\tilde{A}_{1,-1}$,
$\tilde{B}_{11}$, $\tilde{B}_{1,-1}$, $C_{11}$, $C_{1,-1}$ and
the linear relations (\ref{eq:A_lin_1})\textendash (\ref{eq:B_lin_2}),
we obtain

\begin{alignat}{1}
\tilde{A}_{11} & =k_{1}k_{2}\left(1+\frac{\kappa^{2}}{k_{1}^{2}}\right)\left(1+\frac{\kappa^{2}}{k_{2}^{2}}\right)\frac{\left[\left(k_{1}+k_{2}\right)^{2}+\kappa^{2}\right]\left[\left(k_{1}+k_{2}\right)\left(\omega_{1}\omega_{2}+\Delta^{2}k_{1}k_{2}\right)+\left(\omega_{1}+\omega_{2}\right)\left(\omega_{1}k_{2}+\omega_{2}k_{1}\right)\right]}{2\left\{ \left[\left(\omega_{1}+\omega_{2}\right)^{2}-\Delta^{2}\left(k_{1}+k_{2}\right)^{2}\right]\left[\left(k_{1}+k_{2}\right)^{2}+\kappa^{2}\right]-\left(k_{1}+k_{2}\right)^{2}\right\} }C_{10}C_{01},\\
\tilde{A}_{1,-1} & =k_{1}k_{2}\left(1+\frac{\kappa^{2}}{k_{1}^{2}}\right)\left(1+\frac{\kappa^{2}}{k_{2}^{2}}\right)\frac{\left[\left(k_{1}-k_{2}\right)^{2}+\kappa^{2}\right]\left[\left(k_{1}-k_{2}\right)\left(\omega_{1}\omega_{2}+\Delta^{2}k_{1}k_{2}\right)+\left(\omega_{1}-\omega_{2}\right)\left(\omega_{1}k_{2}+\omega_{2}k_{1}\right)\right]}{2\left\{ \left[\left(\omega_{1}-\omega_{2}\right)^{2}-\Delta^{2}\left(k_{1}-k_{2}\right)^{2}\right]\left[\left(k_{1}-k_{2}\right)^{2}+\kappa^{2}\right]-\left(k_{1}-k_{2}\right)^{2}\right\} }C_{10}C_{01},
\end{alignat}

\begin{alignat}{1}
\tilde{B}_{11} & =k_{1}k_{2}\left(1+\frac{\kappa^{2}}{k_{1}^{2}}\right)\left(1+\frac{\kappa^{2}}{k_{2}^{2}}\right)\frac{\left(\omega_{1}+\omega_{2}\right)\left[\left(k_{1}+k_{2}\right)^{2}+\kappa^{2}\right]\left(\omega_{1}\omega_{2}+\Delta^{2}k_{1}k_{2}\right)}{2\left\{ \left[\left(\omega_{1}+\omega_{2}\right)^{2}-\Delta^{2}\left(k_{1}+k_{2}\right)^{2}\right]\left[\left(k_{1}+k_{2}\right)^{2}+\kappa^{2}\right]-\left(k_{1}+k_{2}\right)^{2}\right\} }C_{10}C_{01}\nonumber \\
 & +k_{1}k_{2}\left(1+\frac{\kappa^{2}}{k_{1}^{2}}\right)\left(1+\frac{\kappa^{2}}{k_{2}^{2}}\right)\frac{\left(k_{1}+k_{2}\right)\left\{ 1+\Delta^{2}\left[\left(k_{1}+k_{2}\right)^{2}+\kappa^{2}\right]\right\} \left(\omega_{1}k_{2}+\omega_{2}k_{1}\right)}{2\left\{ \left[\left(\omega_{1}+\omega_{2}\right)^{2}-\Delta^{2}\left(k_{1}+k_{2}\right)^{2}\right]\left[\left(k_{1}+k_{2}\right)^{2}+\kappa^{2}\right]-\left(k_{1}+k_{2}\right)^{2}\right\} }C_{10}C_{01},\\
\tilde{B}_{1,-1} & =k_{1}k_{2}\left(1+\frac{\kappa^{2}}{k_{1}^{2}}\right)\left(1+\frac{\kappa^{2}}{k_{2}^{2}}\right)\frac{\left(\omega_{1}-\omega_{2}\right)\left[\left(k_{1}-k_{2}\right)^{2}+\kappa^{2}\right]\left(\omega_{1}\omega_{2}+\Delta^{2}k_{1}k_{2}\right)}{2\left\{ \left[\left(\omega_{1}-\omega_{2}\right)^{2}-\Delta^{2}\left(k_{1}-k_{2}\right)^{2}\right]\left[\left(k_{1}-k_{2}\right)^{2}+\kappa^{2}\right]-\left(k_{1}-k_{2}\right)^{2}\right\} }C_{10}C_{01}\nonumber \\
 & +k_{1}k_{2}\left(1+\frac{\kappa^{2}}{k_{2}^{2}}\right)\left(1+\frac{\kappa^{2}}{k_{1}^{2}}\right)\frac{\left(k_{1}-k_{2}\right)\left\{ 1+\Delta^{2}\left[\left(k_{1}-k_{2}\right)^{2}+\kappa^{2}\right]\right\} \left(\omega_{1}k_{2}+\omega_{2}k_{1}\right)}{2\left\{ \left[\left(\omega_{1}-\omega_{2}\right)^{2}-\Delta^{2}\left(k_{1}-k_{2}\right)^{2}\right]\left[\left(k_{1}-k_{2}\right)^{2}+\kappa^{2}\right]-\left(k_{1}-k_{2}\right)^{2}\right\} }C_{10}C_{01},
\end{alignat}

\begin{alignat}{1}
C_{11} & =-k_{1}k_{2}\left(1+\frac{\kappa^{2}}{k_{1}^{2}}\right)\left(1+\frac{\kappa^{2}}{k_{2}^{2}}\right)\frac{\left(k_{1}+k_{2}\right)\left[\left(k_{1}+k_{2}\right)\left(\omega_{1}\omega_{2}+\Delta^{2}k_{1}k_{2}\right)+\left(\omega_{1}+\omega_{2}\right)\left(\omega_{1}k_{2}+\omega_{2}k_{1}\right)\right]}{2\left\{ \left[\left(\omega_{1}+\omega_{2}\right)^{2}-\Delta^{2}\left(k_{1}+k_{2}\right)^{2}\right]\left[\left(k_{1}+k_{2}\right)^{2}+\kappa^{2}\right]-\left(k_{1}+k_{2}\right)^{2}\right\} }C_{10}C_{01},\\
C_{1,-1} & =-k_{1}k_{2}\left(1+\frac{\kappa^{2}}{k_{1}^{2}}\right)\left(1+\frac{\kappa^{2}}{k_{2}^{2}}\right)\frac{\left(k_{1}-k_{2}\right)\left[\left(k_{1}-k_{2}\right)\left(\omega_{1}\omega_{2}+\Delta^{2}k_{1}k_{2}\right)+\left(\omega_{1}-\omega_{2}\right)\left(\omega_{1}k_{2}+\omega_{2}k_{1}\right)\right]}{2\left\{ \left[\left(\omega_{1}-\omega_{2}\right)^{2}-\Delta^{2}\left(k_{1}-k_{2}\right)^{2}\right]\left[\left(k_{1}-k_{2}\right)^{2}+\kappa^{2}\right]-\left(k_{1}-k_{2}\right)^{2}\right\} }C_{10}C_{01}.\label{eq:C1-1_final}
\end{alignat}

\end{widetext}

\bibliographystyle{apsrev4-2}

\end{document}